\documentclass[useAMS]{mn2e}

\usepackage{amssymb,amsmath,psfig,times}
\usepackage{graphicx}
\def\gsim{ \lower .75ex \hbox{$\sim$} \llap{\raise .27ex \hbox{$>$}} }
\def\lsim{ \lower .75ex\hbox{$\sim$} \llap{\raise .27ex \hbox{$<$}} }

\def\sc{Schwarzschild}
\def\beq{\begin{equation}}
\def\eeq{\end{equation}}

\def\sc{Schwarzschild}

\def\sw{{\it Swift}}
\def\fe{{\it Fermi}}

\def\ep{$E_{\rm peak}$}

\def\epo{$E^{\rm obs}_{\rm peak}$}

\def\liso{$L_{\rm iso}$}
\def\eiso{$E_{\rm iso}$}
\def\lisocom{$L'_{\rm iso}$}
\def\eisocom{$E'_{\rm iso}$}
\def\ama{$E_{\rm peak}-E_{\rm iso}$}
\def\ghi{$E_{\rm peak}-E_{\gamma}$}
\def\yone{$E_{\rm peak}-L_{\rm iso}$}
\def\amacom{$E'_{\rm peak}-E'_{\rm iso}$}
\def\yonecom{$E'_{\rm peak}-L'_{\rm iso}$}
\def\epcom{$E'_{\rm peak}$}

\def\GA{$\Gamma$}
\def\G{$\Gamma_{0}$}

\title{Gamma Ray Bursts in the comoving frame}

\author[G. Ghirlanda et al.]{G. Ghirlanda
$^{1}$\thanks{E-mail:giancarlo.ghirlanda@brera.inaf.it},
L. Nava$^{2}$, G. Ghisellini$^{1}$, A. Celotti$^{2}$, D. Burlon$^{3}$, S. Covino$^{1}$, A. Melandri$^{1}$ \\
$^{1}$INAF -- Osservatorio Astronomico di Brera, Via E. Bianchi 46, I-23807 Merate, Italy\\
$^{2}$SISSA -- via Bonomea, 265, I-34136 Trieste, Italy\\
$^{3}$Max-Planck-Institut f\"{u}r Extraterrestrische Physik, Giessenbachstra\ss e 1, D-85478 Garching, Germany
}
\begin{document}

\date{}


\maketitle

\label{firstpage}

\begin{abstract}
We estimate the bulk Lorentz factor \G\ of  31 GRBs using the measured peak time of their afterglow light curves. 
We consider two possible scenarios for the estimate of \G: the case of a homogeneous circumburst medium or a wind density profile. 
The values of \G\ are broadly distributed between few tens and several hundreds with average values 
 $\sim$138 and $\sim$66 for the homogeneous and wind density profile, respectively. 
We find that the isotropic energy and luminosity correlate in a similar way with \G, i.e. \eiso$\propto$\G$^2$ 
and \liso$\propto$\G$^2$, while the peak energy \ep$\propto$\G. These correlations are less scattered in the wind 
density profile than in the homogeneous case. We then study the energetics, luminosities and spectral properties of our bursts in their 
comoving frame. The distribution of \lisocom\ is very narrow with a dispersion of less than a decade in the wind case,
clustering around  \lisocom$\sim5\times10^{48}$ erg s$^{-1}$.
Peak photon energies cluster around \epcom$\sim$ 6 keV.
The newly found correlations involving \G\ offer a general interpretation scheme for the 
spectral--energy correlation of GRBs. The \ama\ and \yone\ correlations are due to the 
different \G\ factors and the collimation--corrected correlation, \ghi\ (obtained by correcting 
the isotropic quantities for the jet opening angle $\theta_{\rm j}$), can be explained  
if $\theta_{\rm j}^2\Gamma_{0}=$ constant. 
Assuming the \ghi\ correlation as valid, we find a typical value of 
$\theta_{\rm j}$\G$\sim$  6--20, in agreement with the predictions of 
magnetically accelerated jet models.
\end{abstract}


\section{Introduction}

The discovery of the afterglows of Gamma Ray Bursts (GRBs - Costa et al. 1997) 
allowed to pinpoint their position in the X--ray and Optical bands. This 
opened a new era focused at measuring the spectroscopic redshifts of these sources. 
The present\footnote{http://www.mpe.mpg.de/$\sim$jcg/grbgen.html} collection of GRBs 
with measured $z$ consists of 232 events. In 132 bursts of this sample (updated in this paper) 
the peak energy \epo\ of their $\nu F_{\nu}$ prompt emission $\gamma$--ray spectrum could be constrained. 
In turn, for these bursts it was possible to calculate the isotropic equivalent 
energy \eiso\ and luminosity \liso. The knowledge of the redshifts showed that two 
strong correlations exist between the {\it rest frame} peak energy \ep\ and \eiso\ or \liso\ 
(also known as the ''Amati'' and ''Yonetoku" correlations -- Amati et al. 2002, Yonetoku et al. 2004, respectively). 

The reality of these correlations has been widely discussed in the literature.
Some authors pointed out that they can be the result of observational selection effects 
(Nakar \& Piran 2005; Band \& Preece 2005; Butler et al. 2007, 
Butler, Kocevski \& Bloom 2009; Shahmoradi \& Nemiroff 2011)
but counter--arguments have been put forward arguing that selection effects, 
even if surely present, play a marginal role 
(Ghirlanda et al. 2005, Bosnjak et al. 2008, Ghirlanda et al. 2008; 
Nava et al., 2008; Krimm et al. 2009; Amati et al. 2009).
The finding that a correlation $E_{\rm p}(t)$--$L_{\rm iso}(t)$ exists when
studying time--resolved spectra of individual bursts is a strong argument in favor
of the reality of the spectral energy correlations,  
(Ghirlanda, Nava\& Ghisellini 2010; Ghirlanda et al. 2011) 
and motivates the search for the underlying process generating them.
Even if several ideas have been already discussed in the literature,
there is no general consensus yet, and a step forward towards a better
understanding both of the spectral energy correlations and the
underlying radiation process of the prompt emission of GRBs is
to discover what are the typical energetics, peak frequencies and
peak luminosities in the {\it comoving frame}.

The physical model of GRBs requires that the plasma emitting $\gamma$--rays should be 
moving relativistically with a bulk Lorentz factor \G\ much larger than unity. 
The high photon densities and the short timescale variability of the prompt emission 
imply that GRBs are optically thick to pair production which, in turn, would lead to a 
strong suppression of the emitted flux, contrary to what observed. 
The solution of this compactness problem requires that GRBs are relativistic 
sources. From this argument lower limits $\Gamma_{0} \ge 100$ 
are usually derived (Lithwick \& Sari, 2001). The first observational evidences 
supporting this scenario were found in the radio band where the ceasing of the radio 
flux scintillation (few weeks after the explosion as in GRB 970508; Frail et al. 1997),
allowed to estimate \GA\ of a few. 
This value corresponds to the late afterglow phase, when the fireball is decelerated 
almost completely by the interstellar medium and is characterized
by a much smaller bulk Lorentz factor than the 
typical \G\ of the prompt phase. 

Large Lorentz factors imply strong beaming of the radiation we see. 
We are used to consider GRB intrinsic properties (\ep, \eiso, \liso) for the bursts 
with measured redshifts, but still an important correction should be applied.  
Our aim is to study the distributions of \ep, \eiso, \liso\ and the spectral--energy 
correlations (\ama\ and \yone) in the {\it comoving frame}, accounting for the \G\ factor. 
The estimate of \G\ is possible by measuring the peak of the afterglow (Sari \& Piran 1999) 
and has been successfully applied in some cases (e.g. Molinari et al. 2007, Gruber et al., 2011) 
and more extensively recently by Liang et al. (2010) in the optical and X--ray band. 
Other methods allow to set lower limits (Abdo et al. 2009; Ackerman et al. 2010; 
Abdo et al. 2009a) mainly by applying the compactness argument to the high energy emission 
recently detected in few GRBs at GeV energies by the \fe\ satellite
(see Zou, Fan \& Piran 2011; Zhao, Li \& Bai 2011; Hascoet et al. 2011 for more updated
calculation on these lower limits on \G).
Conversely, upper limits (Zou \& Piran 2010) can be derived by requiring that the forward shock 
emission of the afterglow does not appear in the MeV energy band. 

The paper is organized as follows: in \S~2 we discuss the relativistic corrections that 
allow us to derive the comoving frame \epcom, \eisocom and \lisocom\ from the rest frame 
\ep, \eiso, \liso; in \S~3 and \S~4 we derive a general formula for the estimate of \G\ from the 
measurement of the time of the peak of the afterglow emission; in \S~5 we present our 
sample of GRBs and in \S~6 our results which are finally discussed in \S~7. 
Throughout the paper we assume a standard cosmology 
with $h=\Omega_{\Lambda}=0.7$ and $\Omega_{m}=0.3$.

\section{From the rest to the comoving frame}

In this section we derive the Lorentz transformations to pass from 
rest frame quantities to the same quantities in the comoving frame.
This is not trivial, since, differently from the analog case of blazars,
the emitting region is not a blob with a mono--directional velocity, 
but a fireball with a radial distribution of velocities.
Therefore, an observer located on axis receives photons from a range of viewing angles,
complicating the transformations from rest frame to comoving quantities.
We are interested to three observables: the peak energy $E_{\rm peak}$,
the isotropic equivalent energy $E_{\rm iso}$ and the isotropic equivalent 
peak luminosity $L_{\rm iso}$.
Dealing with isotropic equivalent quantities, we can assume that the emitting region 
is a spherical shell with velocities directed radially.
We also assume that the comoving frame bolometric intensity $I^\prime$ is isotropic.
We then adopt the usual relation between observed ($I$) and comoving ($I^\prime$)
bolometric intensity:
\begin{equation}
I\, =\, \delta^4 I^\prime;\qquad\quad \delta ={1\over \Gamma(1-\beta\cos\theta) }
\end{equation}
where $\delta$ is the Doppler factor and $\theta$ is the angle between the velocity vector and 
the line of sight.
The received flux is
\begin{equation}
F\, =\, 2\pi I^\prime \int_0^{\pi} \delta^4 \sin \theta d\theta 
\end{equation}
Since the fluence ${\cal F}$ is a time--integrated quantity we have 
${\cal F} \propto \int_0^{\pi} \delta^3 \sin \theta d\theta$,
i.e. one power of $\delta$ less.

\vskip 0.2 cm
\noindent
$E_{\rm peak}$ --- This quantity can be derived from the time--integrated spectrum, 
or can be the spectral peak energy of a given time interval.
In this paper we will use the time--integrated $E_{\rm peak}=E_{\rm peak}^{\rm obs}(1+z)$.
The received fluence $d{\cal F}/d\theta$ (i.e. the flux integrated in time) 
from each annulus of same viewing angle 
$\theta$ is $d{\cal F}/d\theta \propto  \sin\theta \delta^3$.
For $\theta\to 0$ the Doppler factor is maximum, but the solid angle vanishes, 
while for $\theta > 1/\Gamma$ the solid angle is large, but $\delta$ is small.
Therefore there will be a specific angle $\theta$ for which $d{\cal F}/d\theta$ is maximum.
This is given by
\begin{equation}
\cos\theta = \beta+ {2\over 5\Gamma^2}
\end{equation}
At this angle the beaming factor is
\begin{equation}
\delta = {5\over 3} \Gamma
\end{equation}
We then set $E^\prime_{\rm peak}= E_{\rm peak}/(5\Gamma/3)$.

\vskip 0.2 cm
\noindent
$E_{\rm iso}$ --- 
This is proportional to the fluence ${\cal F}$, and the relation between the 
observed and comoving quantity is
\begin{equation}
{E_{\rm iso} \over E^\prime_{\rm iso}} =  { {\cal F}\over {\cal F}^\prime}=
{ \int_0^{\pi} \delta^3 \sin \theta d\theta \over
\int_0^{\pi}  \sin \theta d\theta } = \Gamma
\end{equation}
We then set $E^\prime_{\rm iso} = E_{\rm iso}/\Gamma$.

\vskip 0.2 cm
\noindent
$L_{\rm iso}$ --- This is proportional to the flux $F$, so the 
ratio $L_{\rm iso}/L^\prime_{\rm iso}$ is
\begin{equation}
{L_{\rm iso} \over L^\prime_{\rm iso}} =  
{ F\over F^\prime}=
{ \int_0^{\pi} \delta^4 \sin \theta d\theta \over
\int_0^{\pi}  \sin \theta d\theta } \sim {4\over 3} \Gamma^2 
\label{eqlum}
\end{equation}
We then set $L^\prime_{\rm iso} = L_{\rm iso}/(4\Gamma^2/3)$ 
(in agreement with Wijers \& Galama 1999).

\section{Estimate of the Bulk Lorentz Factor \G}


In the thin--shell regime (i.e. for $T_{\rm 90}<t_{\rm peak, obs}$, condition satisfied 
for almost all bursts in our sample) the standard afterglow theory predicts that the peak of the 
bolometric afterglow light curve corresponds to the start of the fireball deceleration. 
The deceleration radius is commonly defined as the radius at which the swept up matter $m(r_{\rm dec})$ is 
smaller by a factor $\Gamma_0$ than the initial shell's rest mass $M_0=E_0/(\Gamma_0c^2)$. 
Usually, the deceleration time $t_{\rm dec}$ is estimated as 
$t_{\rm dec}=r_{\rm dec}/(2c\Gamma_0^2)$ (Sari \& Piran 1999). 
This relation is approximate, since it does not consider that the 
Lorentz factor is decreasing (e.g. Bianco \& Ruffini 2005). 
Some authors consider this relation to estimate $\Gamma_0$ 
from the peak time of the afterglow light curve (Sari \& Piran 1999; Sari 1997), 
while other authors consider that $t_{\rm dec}=r_{\rm dec}/(2c\Gamma_{\rm dec}^2)$, 
where approximately $\Gamma_0\simeq2\Gamma(r_{\rm dec})$ (Molinari et al. 2007).

We propose here a detailed and general calculation of $\Gamma_0$ which extends 
the estimate to the generic case of a circumburst density profile described by 
$n=n_0r^{-s}$. 
We use the shape of the light curve in two different power--law regimes: 
the coasting phase when $r\ll r_{\rm dec}$ and $\Gamma(r)=\Gamma_0$, 
and the deceleration phase when $r_{\rm dec}\ll r \ll r_{\rm NR}$ 
(where $r_{\rm NR}$ marks the start of the non--relativistic regime). 
During the deceleration regime the evolution of the Lorentz factor is described by the 
self--similar solution found by Blandford \& McKee (1976):
\begin{equation}
\Gamma=\sqrt{\frac{(17-4s)E_0}{(12-4s)m(r)c^2}}
\label{bm}
\end{equation}
The relation between the radius and the observed time is obtained by integrating 
the differential equation $dr=2c\Gamma^2(r)dt$ and by considering the exact 
evolution of $\Gamma$ with $r$. 
%
From Eq. \ref{eqlum}:
\begin{equation}
L_{\rm iso}=\frac{4}{3}\Gamma^2L'_{\rm iso}=\varepsilon_e \frac{4}{3}\Gamma^2\frac{dE'_{\rm diss}}{dt'}
\label{eqstart}
\end{equation}
where the dissipated comoving energy $E'_{\rm diss}$ is given by (Panaitescu \& Kumar 2000):
\begin{equation}
E'_{\rm diss}=(\Gamma-1) m(r) c^2
\label{eqediss}
\end{equation}
Only a fraction $\varepsilon_e$ of the dissipated energy is radiated. We assume that 
this quantity is small and does not affect the dynamics of the fireball (adiabatic regime).  
Eq. \ref{eqstart} holds until the emission process is efficient (fast cooling regime).

During the coasting phase $\Gamma=\Gamma_0\gg 1$ and the luminosity (denoted by $L_{\rm iso,1}$) is:
\begin{equation}
L_{\rm iso,1}=\varepsilon_e \frac{4}{3}\Gamma_0^3c^2\frac{dm(r)}{dt'}=
\varepsilon_e \frac{4}{3}\Gamma_0^4c^34\pi r^{(2-s)}n_0m_{\rm p}
\end{equation}
Since in this phase the Lorentz factor is constant and equal to $\Gamma_0$ the 
relation between the fireball radius and the observed time is $$r=2ct\Gamma_0^2$$
As a function of time, the luminosity is:
\begin{equation}
L_{\rm iso,1}=\varepsilon_e \frac{4}{3}2^{(4-s)}\pi n_0m_{\rm p}c^{(5-s)}\Gamma_0^{8-2s}t^{2-s}
\label{liso1}
\end{equation}
For a homogeneous density medium ($s=0$) the light curve rises as $t^2$. 
The luminosity is instead constant when $s=2$, which corresponds to the stellar wind
density profile.

To derive the luminosity during the deceleration phase we start again from 
Eq. \ref{eqstart} and Eq. \ref{eqediss}. However, in this case $\Gamma$ is decreasing 
according to Eq. \ref{bm} (but still $\Gamma\gg1$). We derive:
\begin{equation}
L_{\rm iso,2}=\varepsilon_e \frac{4}{3}\Gamma^2c^2\left[\Gamma\frac{dm(r)}{dt'}+m(r)\frac{d\Gamma}{dt'}\right]
\end{equation}
The first term of the sum in square brackets can be written as 
$$\Gamma\frac{dm(r)}{dr}\frac{dr}{dt'}=(3-s)\frac{m(r)}{r}\Gamma^2c$$
The second term of the sum becomes 
$$m(r)\frac{d\Gamma}{dr}\frac{dr}{dt'}=-\frac{3-s}{2}\frac{m(r)}{r}\Gamma^2c$$

\noindent During the deceleration
$$t=\frac{1}{2c}\int\frac{dr}{\Gamma^2}=\frac{r}{2(4-s)c\Gamma^2}$$ 
where we have used $\Gamma(r)$ given in Eq. \ref{bm}. 

For $\Gamma_0\gg1$ the initial energy content of the fireball $E_0=E_{\rm k,iso}+M_0c^2\simeq E_{\rm k,iso}$, 
where $E_{\rm k,iso}$ is the  isotropic kinetic energy powering the expansion of the fireball in the ISM 
during the afterglow phase.
If the radiative efficiency $\eta$ of the prompt phase is small, $E_{\rm k,iso}$ can be estimated 
from the energetics of the prompt as $E_{\rm k,iso}=E_{\rm iso}/\eta$.  
We obtain:
\begin{eqnarray}
L_{\rm iso,2}&=&\varepsilon_e \frac{4}{3}\Gamma^2c^2\frac{(3-s)m(r)}{4(4-s)t}\\
&=&\varepsilon_e \frac{4}{3}\frac{(17-4s)(3-s)E_{\rm iso}}{4(12-4s)(4-s)\eta}t^{-1}\nonumber
\end{eqnarray}

The peak time of the light curve is the time when the coasting phase ends and 
the deceleration phase starts and can be estimated by setting 
$L_{\rm iso,1} (t_{\rm peak})=L_{\rm iso,2} (t_{\rm peak})$:
\begin{equation}
t_{\rm peak}=\left[\frac{(17-4s)(3-s)E_{\rm iso}}{2^{6-s}\pi 
n_0m_{\rm p}c^{5-s}\eta(12-4s)(4-s)\Gamma_0^{8-2s}}\right]^\frac{1}{3-s}
\end{equation}
and inverting this relation to obtain the initial Lorentz factor as a function of the peak time:
\begin{equation}
\Gamma_0=\left[\frac{(17-4s)(3-s)E_{\rm iso}}{2^{6-s}\pi 
n_0m_{\rm p}c^{5-s}\eta(12-4s)(4-s)t_{\rm peak}^{3-s}}\right]^\frac{1}{8-2s}
\label{eqgamma}
\end{equation}
where $t_{\rm peak}$ is the peak of the afterglow light curve in the source rest frame, i.e. 
$t_{\rm peak}=t_{\rm peak,obs}/(1+z)$, and it will be indicated as $t_{\rm p,z}$ hereafter.

%


While a wind density profile (hereafter W: wind interstellar medium) is  
expected from a massive star progenitor that undergoes strong wind mass losses 
during the final stages of its life (Chevalier \& Li 1999), it is not possible 
at the present stage to prefer the W to the homogeneous interstellar medium case (H, hereafter). 
We already showed (Nava et al. 2006) that the collimation corrected \ghi\ correlation 
(so called ``Ghirlanda" correlation; Ghirlanda, Ghisellini \& Lazzati 2004) has a 
smaller scatter and a linear slope when computed under the assumption of the W compared to the H case. 
It is, therefore, important to compare the estimates of \G\ and of the comoving frame energetics 
in these two possible scenarios. 
The most extensive study of Liang et al. (2010) estimated \G\ mostly from the peak of the 
afterglow light curve in the optical band and in few cases from a peak in the the X--ray band. 
They considered only the H case and found a strong correlation between \G\ and the 
GRB isotropic equivalent energy \eiso. 

Eq. \ref{liso1} predicts that the afterglow light curve is flat in the coasting
phase, with no peaks in the W density case ($s=2$).
However, this equation neglects pre--acceleration of the circumburst matter due to the 
prompt emission itself, that can have important consequences, as we discuss below.




\begin{table*}
\begin{center}
\begin{tabular}{lllllllll}
\hline
\hline
GRB     &z       & \ep\	         &\eiso\              &      \liso\	   & $t_{\rm p,z}$  &    $\Gamma_{\rm H}$ & $\Gamma_{\rm W}$ &Ref\\
        &         & keV             & erg                &  erg/s           &   s                   &                                       &                                        &   \\
\hline
990123	&  1.60    & 2031$\pm$161  & (2.39$\pm$0.28)E54 & (3.53$\pm$1.23)E53 &  18                              &  312        & 182          & 2\\
030226	& 1.986  &  290$\pm$ 63	&  (6.7$\pm1.2$)E52       &  (8.52$\pm$2.23)E51  &	 4340					&   26		&  19			& 5\\
050820A  &  2.612  & 1325$\pm$277  & (9.75$\pm$0.77)E53 &   (91$\pm$6.8)E51    &  108.17$\pm$4.62      &  142        & 93             & 1\\
050922C	&  2.198  &  417$\pm$118   & (4.53$\pm$0.78)E52 &  (190$\pm$2.3)E51   &  42                              &  138        & 55             & 2\\
060210	&  3.91    &  575$\pm$186   & (4.15$\pm$0.57)E53 & (59.5$\pm$8.0)E51   &  97                              &  133        & 77             & 2\\
060418 	&  1.489  &  572$\pm$114   & (1.28$\pm$0.10)E53 & (18.9$\pm$1.59)E51 &  60.73 $\pm$0.82	 &   137       & 65              & 1\\
060605 	&  3.78    &  490$\pm$251   & (2.83$\pm$0.45)E52 &  (9.5$\pm$1.5)E51    &  83.14 $\pm$2.7         &   101      &  41             & 1\\
060607A  &  3.082  &  575$\pm$200   & (10.9$\pm$1.55)E52 &   (20$\pm$2.7)E51    &  42.89 $\pm$0.62       &   153      & 68              & 1\\
060904B  &  0.703  &  135$\pm$41     & (36.4$\pm$7.43)E50 & (7.38$\pm$1.4)E50   &  271.91$\pm$33.75    &   50        & 18              & 1\\
061007 	&  1.261  &  902$\pm$43     & (8.82$\pm$0.98)E53 & (17.4$\pm$2.45E52  &  34.62 $\pm$0.18       &   215      & 121             & 1\\
061121    &  1.314  & 1289$\pm$153  & (2.61$\pm$0.3)E53   &  (141$\pm$1.5)E51   &  250                            &  88         & 54               & 4\\
070110	&  2.352  &  370$\pm$170   &  (5.5$\pm$1.5)E52    & (45.1$\pm$7.52)E50 &  350                            &   64        & 34              & 4\\
071010B	&  0.947  &  101$\pm$23     & (2.12$\pm$0.36)E52 &   (64$\pm$0.53)E50  &  67                              &  105      & 40              & 2\\
080319C  &  1.95    & 1752$\pm$505  &   (15$\pm$0.79)E52  &  (9.5$\pm$0.12)E52  &  117.38$\pm$3.22	 &  109      & 57              & 1\\
080804	& 2.2		& 810$\pm$45	& (1.15$\pm$0.2)E53    &  (2.69$\pm$0.32)E52  &	 40.5					&   157	&   70		& 5\\
080810 	&  3.35    & 1488$\pm$348  & (3.91$\pm$0.37)E53 & (9.27$\pm$0.87)E52 &  27.02 $\pm$0.26	 &  214       & 105          & 1\\
081203A  &  2.1      & 1541$\pm$757  &  (3.5$\pm$0.3)E53    & (28.1$\pm$1.94)E51 &  118.09$\pm$0.46	 &  121       &  70           & 1\\
090102	& 1.547	& 1148$\pm$143&	(2.2$\pm$0.26)E53  &   (8.7$\pm$0.56)E52	   & 20.3					&   221	&   97		& 5\\	
090618	& 0.54		& 155.5$\pm$11	 & (2.53$\pm$0.25)E53 &   (2.05$\pm$0.1)E52		& 51.9					&   158	&	80		& 5\\
090812	&  2.452  & 2023$\pm$663  & (4.03$\pm$0.4)E53   & (95.6$\pm$9.66)E51 &  17.38	                  &   253    & 118           & 5\\
091024    &  1.092  &  794$\pm$231   &  (2.8$\pm$0.3)E53    & (1.0$\pm$0.22)E52  &  1912	                  &   59   & 66             & 6\\
091029	&  2.752  &  230$\pm$66     &  (7.4$\pm$0.74)E52  & (13.2$\pm$0.73)E51  &  88                             &  111     & 51             & 5\\
100621A  &  0.542  &  146$\pm$23.1  & (4.37$\pm$0.5)E52   & (3.16$\pm$0.24)E51  &  3443	                  &  26       & 18             & 5\\
100728B  &  2.106  &  404$\pm$29     &  (3.0$\pm$0.3)E52    & (18.6$\pm$1.20)E51  &  16                             & 188      & 63             & 5\\
100906A  &  1.727  &  158$\pm$16     & (3.34$\pm$0.3)E53   & (24.5$\pm$0.86)E51  &  37                             & 186      & 93             & 5\\
110205A  &  2.22    &  715$\pm$239   &  (5.6$\pm$0.6)E53    & (2.50$\pm$0.34)E52  &  311                           &  89       & 62             & 5\\
110213A  &  1.46    &  241$\pm$13     &  (6.4$\pm$0.6)E52    & (20.9$\pm$0.58)E51  &  81                             & 113      & 51             & 5\\
\hline
080916C  &  4.35    & 2759$\pm$120  &  (5.6$\pm$0.5)E54    & (10.4$\pm$0.88)E53 &   1.5	                          &  880      & 419           & 3\\
090510	&  0.903  & 4400$\pm$400  &  (5.0$\pm$0.5)E52    & (1.78$\pm$0.12)E53 &  0.44(315.3)	         & 773(66)       & 175(34)          & 3(7)\\
090902B  &  1.822  & 2020$\pm$17    &   (44$\pm$0.3)E53    & (58.9$\pm$0.97)E52 &   3.2	                          & 643      & 327           & 3\\
090926A	&  2.106  &  907$\pm$7	    &   (20$\pm$0.52)E53  &   (74$\pm$1.45)E52  &   2.9	                          & 605      & 275           & 3\\
\hline
\hline
\end{tabular}
\caption{The sample of GRBs with redshifts $z$, rest frame peak energy \ep, 
isotropic equivalent energy \eiso\ and luminosity \liso\ (integrated in the 1 keV--10 MeV energy range) 
and peak time of the  optical afterglow light curve (given in the source rest frame $t_{\rm p,z}$). 
The \G\ factors computed in the H and W case are reported. The GRBs shown separately at the bottom of the table are the three long GRBs 
(080916C, 090902B, 090926A) showing a peak of the GeV light curve (as detected by \fe-LAT) which could be interpreted as afterglow 
emission (Ghisellini et al. 2010). The short GRB 090510 is shown with two entries: one corresponding to the peak of the GeV light curve and the 
second to the peak of the optical light curve. 
The last column gives the references for the peak time of the afterglow: 
(1) Liang et al. 2010, peak of the optical light curve; 
(2) Liang et al. 2010, references in their Tab. 6; 
(3) Ghisellini et al. 2010; e
(4) Ghisellini et al., 2009;
(5) GRBs added in this work (Melandri et al. 2011);
(6) Gruber et al. 2011;
(7) De Pasquale et al. 2009.
}
\label{tab1}
\end{center}
\end{table*}

\section{Homogeneous or wind density profile?}

In the following we will find the initial bulk Lorentz factor $\Gamma_0$
for bursts showing a peak in their early afterglow light curve.
In the simple case of an homogeneous circumburst density,
we expect that the afterglow luminosity $L_{\rm aft} \propto t^2\Gamma^8$, and therefore 
$L_{\rm aft} \propto t^2$ when $\Gamma=\Gamma_0=$ constant
(Eq. \ref{liso1}).
It can be questioned if, in the case of a wind density profile,
such a peak occurs, or if the initial light curve is flat (i.e. $\propto t^0$),
as suggested by Eq. \ref{liso1} when $s=2$.

The derivation leading to Eq. \ref{liso1} assumes that the circumburst
medium is at rest when the fireball impacts through it (i.e. it is an {\it external} shock).
Instead, since the electrons in the vicinity of the burst scatter part of the
prompt emission of the burst itself, some radial momentum has to be transferred
to the medium (as suggested by Beloborodov 2002).
If the velocity acquired by the circumburst matter becomes relativistic, then
the fireball will produce an {\it internal} shock when passing through the medium,
with a reduced efficiency.

To illustrate this point, let consider an electron at some distance $r$ from
the burst, scattering photons of the prompt emission of energy 
$E_{\rm peak} = x m_{\rm e}c^2$. 
In the Thomson limit of the scattering process, this electron will scatter a 
number $\tau$ of prompt photons given by:
\begin{equation}
\tau = \sigma_{\rm T} n_\gamma \Delta r = 
{\sigma_{\rm T} L_{\rm iso} c t_{\rm burst} \over 4\pi r^2 c \, x m_{\rm e} c^2 } 
= { \sigma_{\rm T} E_{\rm iso} \over 4\pi r^2 x m_{\rm e} c^2} 
\end{equation}
To evaluate the distance $r$ up to which this process can be relevant, 
consider at what distance the electrons make a number $\tau \approx (m_{\rm p}/m_{\rm e})/x$
scatterings, namely the distance at which the electrons and their associated protons
are accelerated to $\gamma\sim 2$:
\begin{equation}
r(\gamma=2) \approx \left [ { \sigma_{\rm T} E_{\rm iso} \over 4\pi  m_{\rm p} c^2} \right]^{1/2}
\sim 1.9 \times 10^{15} E_{\rm iso, 53}^{1/2} \quad {\rm cm}
\end{equation}
where $E_{\rm iso, 53}=10^{53} E_{\rm iso}$  erg.
This distance must be compared with the deceleration radius $r_{\rm dec}$
in the case of a wind density profile corresponding to a mass loss 
$\dot M$ and a velocity $v_{\rm w}$ of the wind:
\begin{equation}
n(r)  = {\dot M \over 4\pi r^2 m_{\rm p} v_{\rm w} } =3.16\times 10^{35} 
{\dot M_{\rm -5} \over v_{\rm w, 8} r^2}
\end{equation}
where $\dot M = 10^{-5} \dot M_{-5} M_\odot\, {\rm yr}^{-1}$ and
$v_{\rm w} =10^8$ cm s$^{-1}$ (i.e. $10^3$ km s$^{-1}$) (e.g. Chevalier \& Li 1999).
The deceleration radius is
\begin{equation}
r_{\rm dec}  = {E_{\rm iso} \over 4\pi m_{\rm p} c^2 \eta \Gamma_0^2}
\sim 1.7\times 10^{16} \, {E_{\rm iso, 53} v_{\rm w, 8} \over \eta_{-1} \dot M_{-5} \Gamma^2_{0, 2} }\,\, {\rm cm} 
\end{equation}
where $\eta$ is the efficiency of conversion of the kinetic energy to radiation 
($L_{\rm iso} = \eta L_{\rm k, iso}$).
Therefore it is possible to have a pre--acceleration of the circumburst
matter up to a distance comparable to (but less than) the deceleration radius.
In this case we expect to have a very early {\it rising} afterglow light
curve (corresponding to relatively inefficient internal shocks between the fireball 
and the pre--accelerated circumburst medium), followed by a flat light curve
and then a decay.

We conclude that the absence of a flat early light curve does not
exclude (a priori) a wind density profile.
This gives us a motivation to explore both cases (i.e. homogeneous
and wind density profile) even if the bursts in our sample all
show a peak in the afterglow light curve (and thus a rising phase).

Note that the same pre--acceleration can occur if the density is homogeneous.
In this case, again, we expect the very early afterglow to be less efficient
than what predicted without pre--acceleration, leading to a rising
phase even harder than $t^2$.

\section{The sample}

Since we want to study the energetics, luminosities and peak energies of GRBs in the comoving frame, 
our first requirement is to know the redshift $z$. 
Then we also need that the spectral peak energy \epo\ has been determined from the fit 
of the prompt emission spectrum.
Most of these bursts have been localized by the Burst Alert Telescope 
(BAT; Barthelmy et al. 2005) on board the \sw\ satellite, but only
for a few of them BAT could determine \epo\ (due to its limited energy range, 15--150 keV). 
Most of the \epo\ were determined by the Konus--Wind satellite 
(Aptekar et al. 1995), 
or, since mid 2008, by the Gamma Burst Monitor (GBM; Meegan et al. 2009 with 
energy bandpass 8 keV--35 MeV) on board the \fe\ satellite.
Our sample of GRBs with $z$ and constrained \epo\ (and consequently 
with computed \eiso\ and \liso) is updated up to May 2011.
It contains 132 GRBs with $z$, \epo\ and \eiso.
We have \liso\ for all but one of these bursts.

Within this sample, we searched the literature for bursts with evidence of the peak of the 
afterglow or an estimate of the \G\ factor:
\begin{enumerate}
\item 
Liang et al. (2010 -- L10 hereafter) measured the peaks in the 
optical light curves of GRBs and then estimated \G\ for the H case. From L10 we 
collected 9 measurements of $t_{\rm p,z}$. 
L10 also collected other estimates of $t_{\rm p,z}$ from the literature (their table 6) 
from which we get other 4 values of this observable. 
Therefore from L10
we collected 13 estimates of $t_{\rm p,z}$ from the optical light curves;
\item 
two GRBs, not included in the sample of L10, that show a peak in their 
optical afterglow light curves are taken from Ghisellini et al. (2009);
\item 
L10 searched for bursts with evidence of the afterglow peak up to December 2008. 
Our sample of bursts with redshifts, \epo\ 
and isotropic energies/luminosities extends to May 2011. 
We searched in the literature for $t_{\rm p,z}$ of bursts after 
December 2008 and in  10 cases we could build the light curve with 
available published data (that will be presented in a 
forthcoming paper -- Melandri et al. 2011).  Our systematic search 
of the literature resulted in other 2 GRBs with a peak in the optical light curve.
\end{enumerate}
%


Our sample is thus composed of  
27 GRBs with an estimate of  $t_{\rm p,z}$ obtained from 
their optical light curves. All these are long GRBs. 

The sample is presented in Tab. \ref{tab1} where we show the 
relevant properties of these bursts used in the following sections. Col. 1 and 2 show the GRB name and its redshift, 
Col. 3 the rest frame peak energy \ep, and Col. 4 and 5 the isotropic equivalent energy 
\eiso\ and luminosity \liso, respectively. 
In Col. 6 it is reported the rest frame $t_{\rm p,z}$ from which we compute the 
\G\ factor in the H case (Col. 7) and in the W case (Col. 8) 
assuming a typical density value  $n_0=3$ cm$^{-3}$ or $n_0=3\times 10^{35} {\rm cm^{-1}}$ 
(for the H and W respectively) and a typical radiative efficiency $\eta=0.2$. 
We note from Eq. \ref{eqgamma} that the resulting \G\ is rather insensitive to 
the choice of $n_0$ and $\eta$ both in the H case [i.e. $\Gamma_{0} \propto (n_0 \eta)^{-1/8}$] 
and in the W case [i.e. $\Gamma_{0} \propto (n_0 \eta)^{-1/4}$].

There are also four GRBs, detected by the Large Area Telescope on board \fe\ 
at GeV energies, showing a peak in their GeV light curves 
(Ghisellini et al. 2010). The interpretation of the GeV emission as afterglow (Barniol Duran \& Kumar 2009, 
Ghirlanda et al. 2010, Ghisellini et al. 2010) is however debated (Ackermann et al. 2010; Piran \& Nakar 2010). 
Among these bursts there is also the short/hard GRB 090510 whose \G\ is derived from the modeling of
the GeV light curve (Ghirlanda et al. 2010a). However, this burst also shows a clear peak in the optical at $\sim$300 s after the 
GRB onset  (De Pasquale et al. 2009) which questions the afterglow interpretation 
of the GeV emission.

The three LAT bursts with $t_{\rm p,z}$ measured from the GeV light curve and the short GRB 090510 are 
shown separately in Tab. \ref{tab1}. These events have the smallest $t_{\rm p,z}$ in our sample 
and, therefore, the largest \G\ values (see Tab. \ref{tab1}). This is expected since, as discussed in Ghisellini et al. (2010), the 
detection in the GeV energy range by LAT seems to be a characteristic of GRBs with the largest values of \epo. Besides, 
the possible measure of $t_{\rm p,z}$ in the optical range is limited by the time delay of the follow up of GRBs in this band, although 
several GRBs have been repointed in the optical band by UVOT on board \sw. In the end, there could be a selection bias on the 
bursts with a peak in the GeV energy range, coupled with the debated interpretation of the GeV emission as afterglow. 
For these resons, in the next sections we will present the results of the study of the correlations
between the GRB energetics and \G\ both including and excluding these bursts. In all our quantitative analysis 
we always excluded the short GRB 090510 which is only shown for comparison with the properties of the 27 long GRBs.

In our sample we do not include upper limits on $t_{\rm p,z}$ which are those bursts observed early in the optical 
whose light curve is decaying up to several days without any sign of a peak. Several of these cases can be found in the literature and 
they would provide lower limits on the value of \G. However, it is hard to define an appropriate sample of upper limits on $t_{\rm p,z}$   derived 
from the optical band because of the lack of a unique follow-up program dedicated to the systematic observations of GRB afterglows.

\begin{figure}
\vskip -0.5 cm
\hskip -1.2cm
\psfig{file=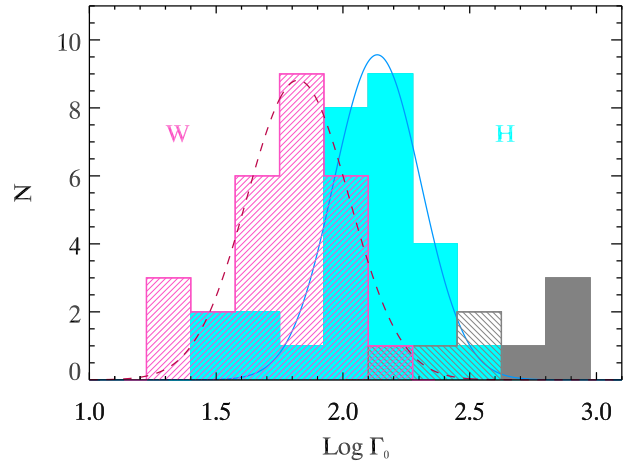,width=10cm}
\vskip -0.2 cm
\caption{
\G\ distributions of the 31 GRBs in the case of an homogeneous interstellar medium (H -- solid filled blue histogram) 
and in the case of a wind density profile (W -- hatched histogram).  The Gaussian functions show the fits (solid and dashed 
line for the H and W case, respectively) to the histograms of the sample of 27 GRBs with   $t_{\rm p,z}$ derived from the optical 
light curve. The three long and one short GRBs with  $t_{\rm p,z}$ measured from the GeV light curve are shown by the grey solid and 
hatched histograms, for the H and W case respectively, but are not included in the fits.
}
\label{fg1}
\end{figure}

\section{Results}

In this section we first show the distributions of the \G\ factors computed 
 in the H and  W and show the correlation of \G\ with the isotropic energy \eiso\ and luminosity \liso. 
Then we show how the distributions of \ep, \eiso\ and \liso\ change when they are corrected for the 
\G\ factor, i.e. how they appear in the comoving frame  (\epcom, \eisocom, \lisocom). 
In doing this we always consider the two estimates of \G\ in the H and W to compare the 
different distributions of the spectral parameters. 
Finally, we present the rest frame \ama\ and \yone\ correlations (updated here with 132 and 131 GRBs 
up to May 2011) and, for those bursts in our sample with measured \G, we show where they cluster in 
these planes when the beaming corrections ($E^\prime_{\rm peak}= E_{\rm peak}/(5\Gamma/3)$,  
$E^\prime_{\rm iso} = E_{\rm iso}/\Gamma$, $L^\prime_{\rm iso} = L_{\rm iso}/(4\Gamma^2/3)$) are applied.

For all the reasons outlined in \S 5, in the following we consider: 
\begin{itemize}
\item the optical sample of 27 GRBs with measured $z$, \epo, \eiso\ and \liso, whose $t_{\rm p,z}$ is measured from 
the optical light curve. 
\item the extended sample of 30 GRBs which includes the three long GRBs with a peak in the GeV which, if interpreted 
as afterglow emission, allows to estimate the largest \G\ in our sample. 
\end{itemize}

\subsection{\G\ distributions}

Fig. \ref{fg1} shows the distributions of the \G\ factors of the  27 
GRBs of our sample (with $t_{\rm p,z}$ measured from the optical light curve - Tab.1)
computed in the H (solid histogram) 
and W case (hatched histogram), respectively. 
The two distributions are fitted with Gaussian functions 
and the central value and dispersion are reported in Tab. \ref{tab2}. 
The average \G\ factor is $\sim$138 in the H case and $\sim$66 in the W case.
In both the H and W case the distribution of \G\ is broad, 
spanning nearly one decade. 

\begin{table}
\begin{center}
  \begin{tabular}{lllll}
    \hline\hline
              & Parameter         & \#GRBs   & Central value & Dispersion ($\sigma$)\\
\hline
    		& $\log E_{\rm peak}$  & 132 &  2.68   & 0.43     \\
		& 				    & 27 &   2.81  &    0.50  \\
    		& 				    & 30 &   2.85  &    0.35  \\
		& $\log E_{\rm iso}$     & 132 &  53.05   & 0.77     \\
		& 				    & 27 &   53.19  &    0.64  \\
		& 			            & 30 &  53.25   & 0.71     \\
		& $\log L_{\rm iso}$     & 131 &  52.46   & 0.73     \\
		& 				   & 27 &  52.53 &    0.82  \\
		& 				   & 30 &  52.62   & 0.87     \\
\hline\hline
    Density &          &   & \\
    \hline
   H & $\log \Gamma_{0}$ & 27 & 2.14 & 0.17  \\
      & 						  & 30 & 2.14 & 0.18  \\ 
    		& $\log E^\prime_{\rm peak}$  & 27 &  0.49   & 0.35     \\
		&								    & 30 &  0.44   & 0.38		\\
    		& $\log E^\prime_{\rm iso}$  & 27 &  51.14   & 0.49     \\
		&								 & 30 &  51.22   & 0.55   \\
		& $\log L^\prime_{\rm iso}$  & 27 &  48.12   & 0.47     \\	
		&								& 30 & 48.11    & 0.39   \\
    \hline
   W & $\log \Gamma_{0} $    & 27 & 1.82 &  0.20  \\  
         & 						  & 30 & 1.82 & 0.21   \\  
    		& $\log E^\prime_{\rm peak} $  & 27 &  0.79   & 0.24     \\
      	& 						  			& 30 & 0.76 & 0.27  \\ 		
    		& $\log E^\prime_{\rm iso} $  & 27 &  51.47   & 0.43     \\
      	& 						  		& 30 & 51.54    & 0.45  \\ 		
		& $\log L^\prime_{\rm iso} $  & 27 &  48.69   & 0.26     \\
      	& 						  		& 30 & 48.71 & 0.23  \\ 
\hline
\end{tabular}
\caption{
Central values and dispersions of the Gaussians fitted to the distributions 
of \G, \ep\ and \epcom, \eiso\ and \eisocom, \liso\ and \lisocom.  
For each quantity we report the Gaussian fits to the sample of 27 GRBs with
$t_{\rm p,z}$ measured from the optical light curve and the sample of 30 GRBs which 
includes the three events with $t_{\rm p,z}$ measured from the GeV light curve, if interpreted
as afterglow. 
The short GRB 090510 has been excluded from this analysis.
} 
\label{tab2}
\end{center}
\end{table}

\begin{center}
\begin{figure*}
\hskip 0.8cm
\psfig{file=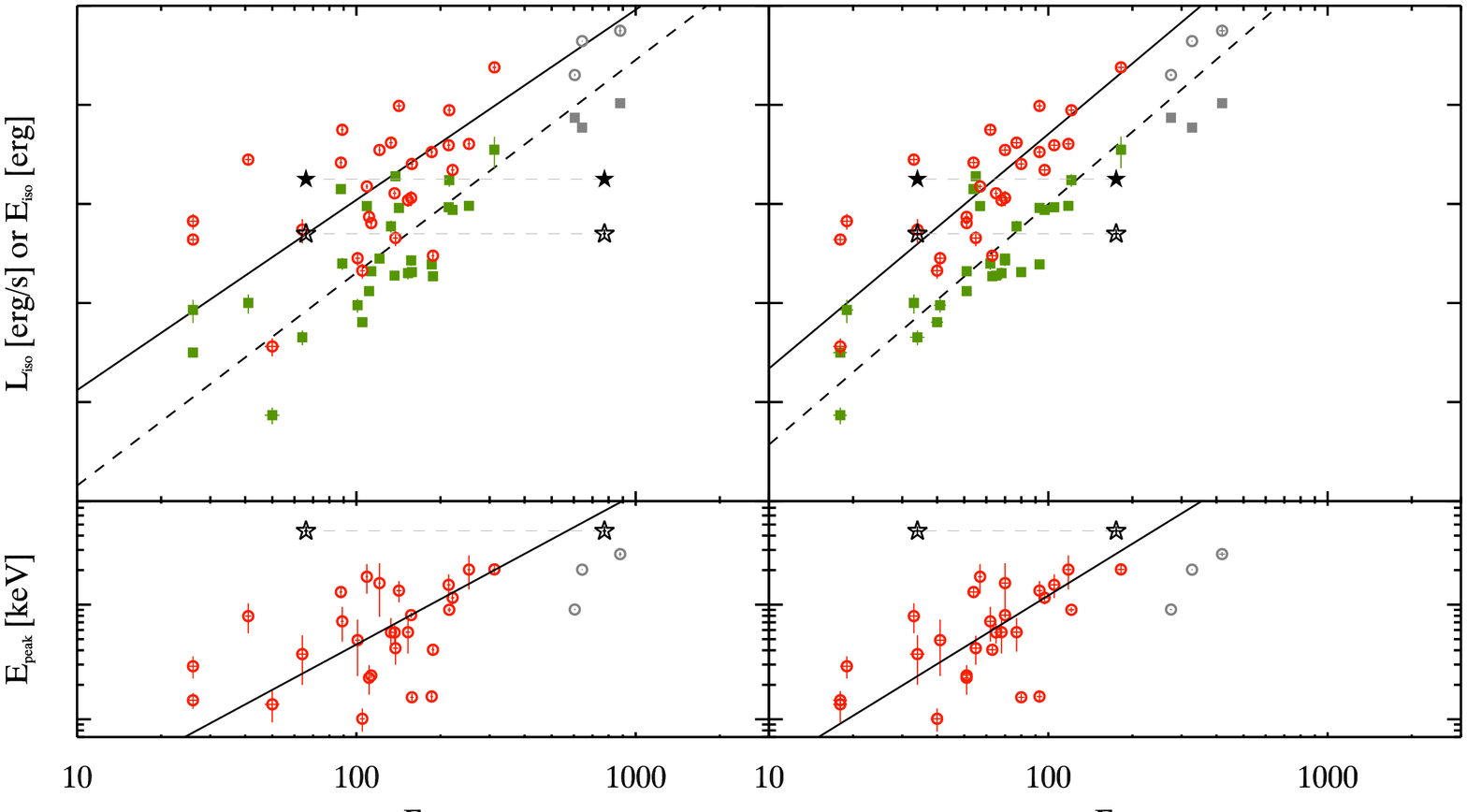,width=17cm}
\vskip 0.5cm
\caption{
{\it Top panels:} Isotropic equivalent energy \eiso\ 
(open circles) and luminosity \liso\ (filled squares) as a function of \G,
computed for the 30 GRBs in our sample in the H case (left panel) and W (right panel). 
The solid (dashed) line in both panels show the least square fit with a 
power law to the \eiso--\G\ (\liso--\G) correlation  
to the sample of 27 GRBs with peak in the optical light curve 
(open red circles and filled green squares). The three GRBs with peak in  the 
GeV light curve are shown with the grey symbols but are not included in the fits shown here. The short GRB 090510
with both a peak in the GeV and a delayed peak in the optical (see Tab. \ref{tab1}) is shown by star symbols connected by
the dashed (gray) line. The larger value of \G\ is that derived from the peak in the GeV light curve.
{\it Bottom panels:} Peak energy \ep\ for the H case 
(left panel) and W case (right panel) as a function of \G.
The solid line is the best fit correlation. 
The correlation coefficient and the slope and normalization of the best 
fit correlations are reported in Tab. \ref{tab3}.
}
\label{fg2}
\end{figure*}
\end{center}

\subsection{\eiso--\G, \liso--\G, \ep--\G\ correlations}

In this section we explore the presence of correlations between the rest frame GRB properties 
(i.e. the peak energy \ep, the isotropic equivalent energy \eiso\ and luminosity \liso) and the \G\ factor. 


In the upper panels of Fig. \ref{fg2} we show the isotropic energy \eiso\ and luminosity \liso\ 
(open red circles and filled green squares, respectively) as a function of
\G\ in both the H and W case (left and right panel, 
respectively).
In the bottom panels of Fig. \ref{fg2} we show the peak energy 
\ep\ as a function of \G\ in the H (left panel) and W (right panel) case. 

The Spearman rank correlation coefficients and associated chance probabilities are reported in 
Tab. \ref{tab3}. 
We model the correlations with a power law: $\log Y =m\, \log \Gamma_{0} +q$ 
(with $Y$=\eiso, $Y$=\liso\ or $Y$=\ep) and list the best fit parameters in Tab. \ref{tab3}. 
We fit this model to the data points (shown in Fig. \ref{fg2}) 
with the bisector method. 
The choice of this fitting method, instead of the least 
square Y vs. X method that minimizes the vertical distances of the data from the fitting 
line, is motivated by the large dispersion of the data and the absence of any physical
motivation for assuming that \G\ or instead \eiso, \liso\ or \ep\ are the independent 
variable (Isobe et al. 1990). 
 
In a  recent work, Lv et al. (2011) derive a correlation 
$\Gamma_{0}\propto E_{\rm iso}^{0.22}$, similar to that found in L10. Such a flat correlation is obtained because
\G\ is fitted versus \eiso\ (or \liso). As described above, the large scatter of the correlations and the lack of any 
physical reason for assuming either \G\ or \eiso\ (\liso) as the independent variable, 
requires instead that these correlations
are fitted with the bisector method. This gives different  correlation slopes with respect to those reported in L10 and Lv et al. (2011). 
Moreover, in our sample we only consider bursts with firm estimates 
of \ep\ and do not include those GRBs which are fitted by a simple power law in the BAT energy range but whose peak energy 
is derived through a Bayesian method, based on the properties of bright BATSE bursts (Butler et al. 2008).

We find that there are strong correlations  between the spectral 
peak energy and isotropic energy/luminosity with \G. 
The slopes of these correlations are rather 
insensitive to the circumburst profile adopted in deriving \G\ (H or W) and are 
similar for \eiso\ and \liso\ (\eiso$\propto$\G$^{2}$ and \liso$\propto$\G$^{2}$). 
A roughly linear correlation exists between \ep\ and \G: \ep$\propto$\G\ (bottom panels in Fig. \ref{fg2}).

The dispersion of the data points  around the best fit correlations 
(shown by the solid and dashed lines in Fig. \ref{fg2}) is modeled with a 
Gaussian and its $\sigma_{\rm sc}$ is given in Tab. \ref{tab3}. 
The less dispersed correlation is between the luminosity \liso\ and \G  
(with $\sigma_{\rm sc}=0.07$).

We finally verified that there is no correlation between the GRB duration $T_{\rm 90}$ and 
\G\ (chance probability $P=0.3$ and $P=0.7$ for the H and W case) and between the 
redshift $z$ and \G.

\begin{figure}
\hskip -1.2cm
\psfig{file=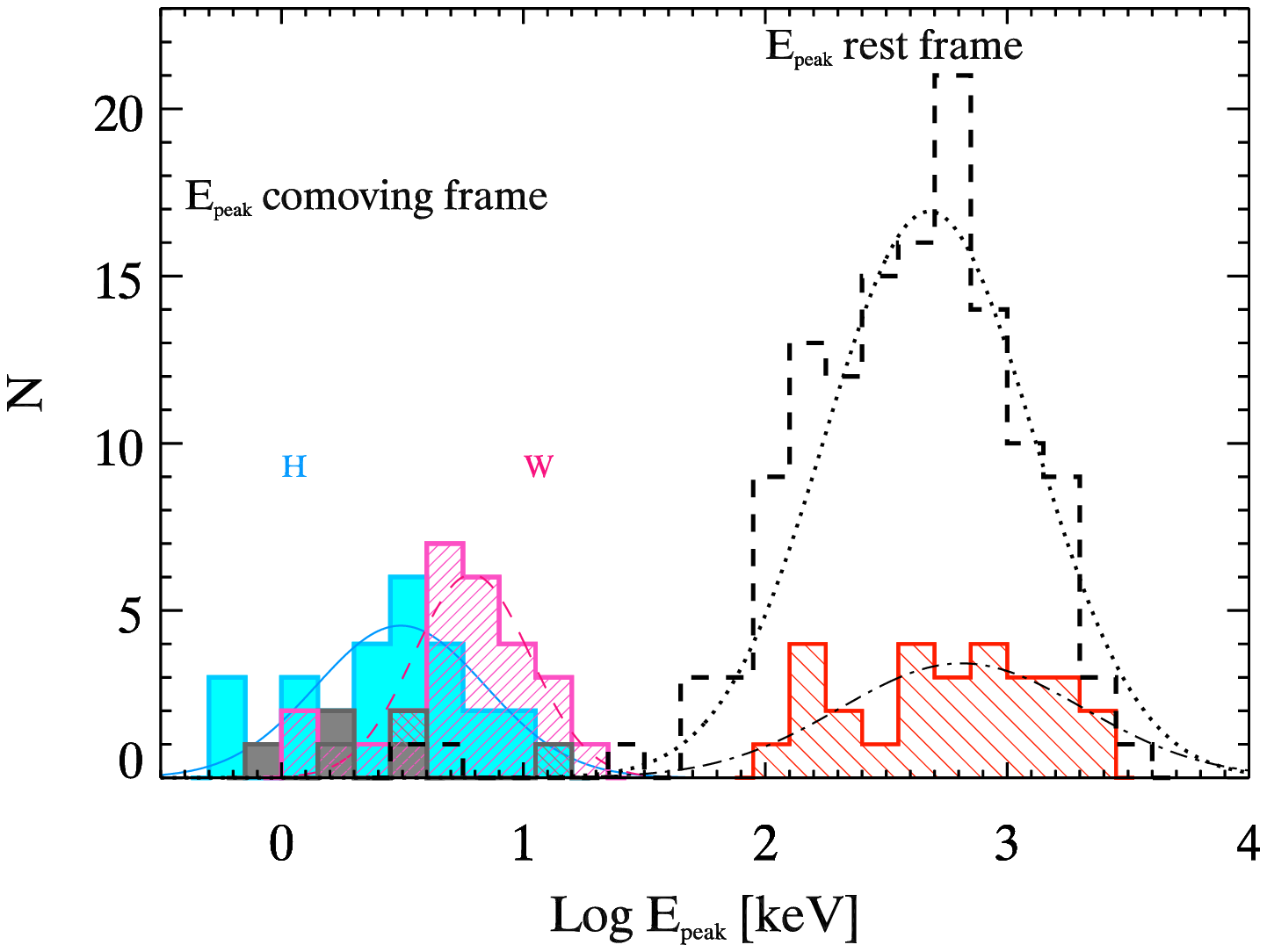,width=10cm}
\caption{
Peak energy distributions in the rest frame \ep\ (dashed histogram) for the sample of 
132 GRBs with known redshift and constrained \ep. 
The hatched histogram shows the 30 GRBs of our sample 
for which we have an estimate of the peak of the afterglow and hence of \G. 
The beaming corrected distribution of \epcom=\ep/(5\G/3) is shown by the solid filled (cyan) 
histogram in the H case and with the hatched (purple) histogram in the W case. For 
all the distributions we also show the  Gaussian fits whose parameters are reported in Tab. \ref{tab2} .  
The four GRBs with a peak in the GeV light curve are shown with gray filled and hatched histograms.
}
\label{fg3}
\end{figure}
\begin{figure}
\hskip -1.2cm
\psfig{file=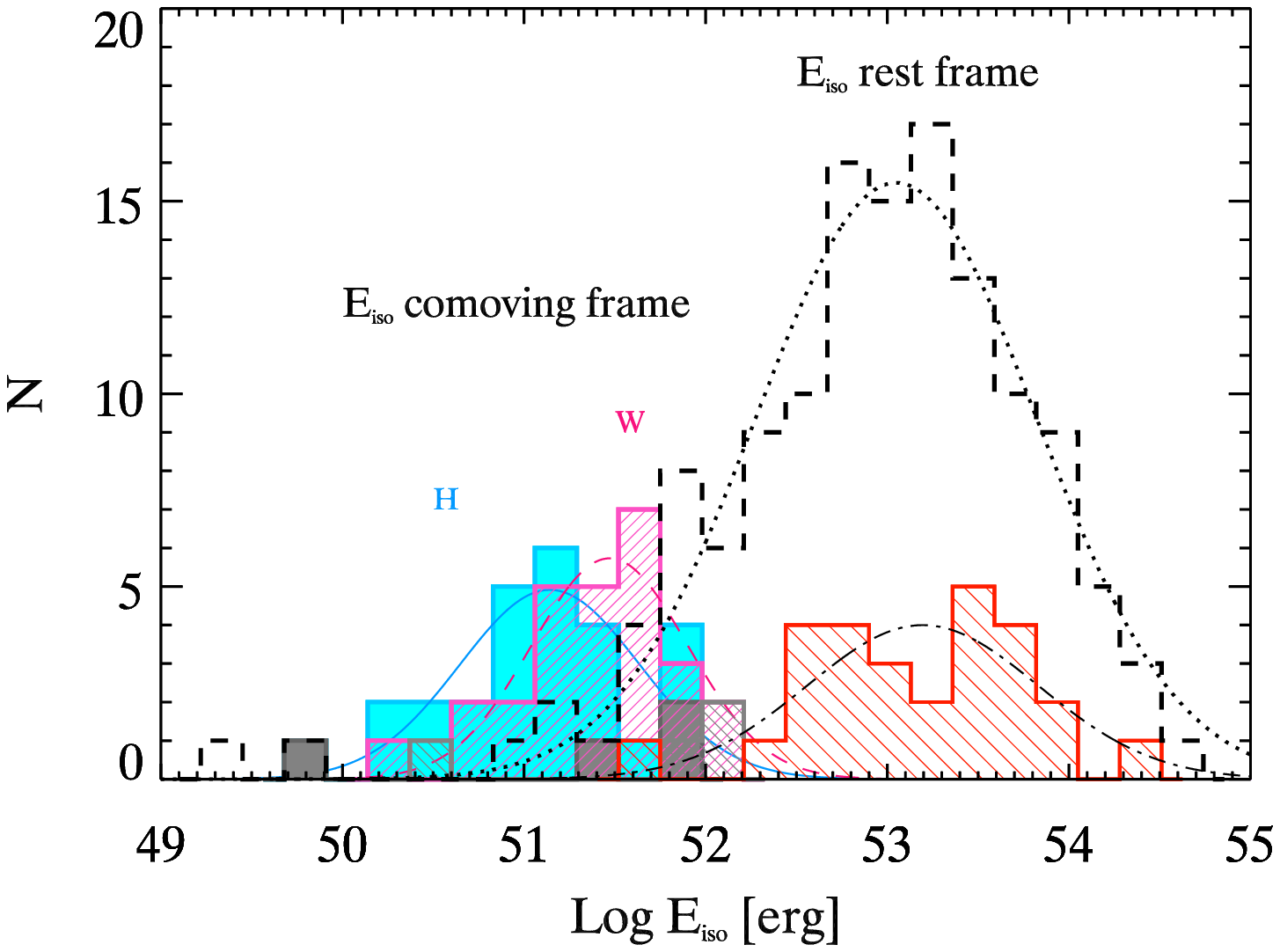,width=10cm}
\caption{
Isotropic energy distributions in the rest frame (dashed histogram) for the sample of 132 GRBs with 
known redshift and constrained \epo. The hatched histogram shows the 30 GRBs of our sample for 
which we have an estimate of the peak of the afterglow. The beaming corrected distribution of 
\eisocom=\eiso/\G is shown by the solid filled histogram and hatched purple histogram for the H and W case 
for the 27 GRBs with a peak in the optical light curve. The four GRBs with a peak in the GeV light curve are shown for 
comparison with the hatched and filled gray histograms.
}
\label{fg4}
\end{figure}

\begin{table*}
\begin{center}
  \begin{tabular}{lllllll}
    \hline\hline
Correlation           & \#GRBs &  $\rho$ & $P_{\rm chance}$ & $m$ & $q$ & $\sigma_{\rm sc}$ \\
\hline
    $E_{\rm iso}-\Gamma_{0}^{\rm H}$ &  27  &0.48  & $10^{-2}$ & 1.92$\pm$0.40 & 49.20$\pm$0.88 & 0.28 \\ 
    											& 30  & 0.74 & $2.5\times10^{-4}$             & 1.96$\pm$0.26 &	49.11$\pm$0.62 & 0.23 \\
    $L_{\rm iso}-\Gamma_{0}^{\rm H}$ &  27  & 0.64 & $3\times10^{-4}$  & 2.15$\pm$0.34 & 48.01$\pm$0.74 & 0.18 \\ 
    											&  30 & 0.74 & $3\times10^{-6}$		  & 2.04$\pm$0.22 &	48.21$\pm$0.51 & 0.20  \\
    $E_{\rm peak}-\Gamma_{0}^{\rm H}$ & 27	& 0.45 & $10^{-2}$ & 1.31$\pm$0.2 & 0.03$\pm$0.36 & 0.21 \\ 
        									        & 30	& 0.56 & $10^{-3}$ & 1.13$\pm$0.13 & 0.36$\pm$0.31 & 0.23 \\
\hline
    $E_{\rm iso}-\Gamma_{0}^{\rm W}$ & 	27	&	0.75 & $4\times10^{-4}$            & 2.36$\pm$0.36 & 48.97$\pm$0.60 & 0.18 \\ 
        										&  30	& 0.82 & $2.2\times10^{-8}$  & 2.15$\pm$0.20 & 49.32$\pm$0.42 & 0.10 \\
    $L_{\rm iso}-\Gamma_{0}^{\rm W}$ & 	27	&   0.76 & $5\times10^{-6}$ & 2.40$\pm$0.24 & 48.14$\pm$0.43 & 0.07 \\ 
        									     &  30   & 	0.82  & $2.6\times10^{-8}$ & 2.19$\pm$0.16 & 48.52$\pm$0.31 & 0.10 \\
    $E_{\rm peak}-\Gamma_{0}^{\rm W}$ & 27	&   0.62 & $5\times10^{-4}$    & 1.50$\pm$0.20 & 0.08$\pm$0.30 & 0.25 \\ 
        										   &  30& 	0.69 &	$2.3\times10^{-5}$ & 1.21$\pm$0.20 & 0.54$\pm$0.27 & 0.31 \\
\hline
\end{tabular}
\caption{
Results of the fit of the \G--\eiso, \G--\liso\ and \G--\ep\ correlations in the 
two cases of homogeneous insterstellar medium (H) and wind density profile (W). 
The Spearman correlation coefficient $\rho$ and the chance probability $P_{\rm chance}$ 
are reported together with the slope $m$ and normalization $q$ 
of the fit of the data points with a linear model 
The fit is done with the bisector method  considering the sample of 27 GRBs with optical peak and the 30 GRBs (i.e. including the three long bursts with peak in the GeV).
} 
\label{tab3}
\end{center}
\end{table*}

\subsection{Comoving frame \epcom, \eisocom, \lisocom\ distributions}

\begin{figure*}
\vskip -0.5 cm
\hskip -1.2cm
\psfig{file=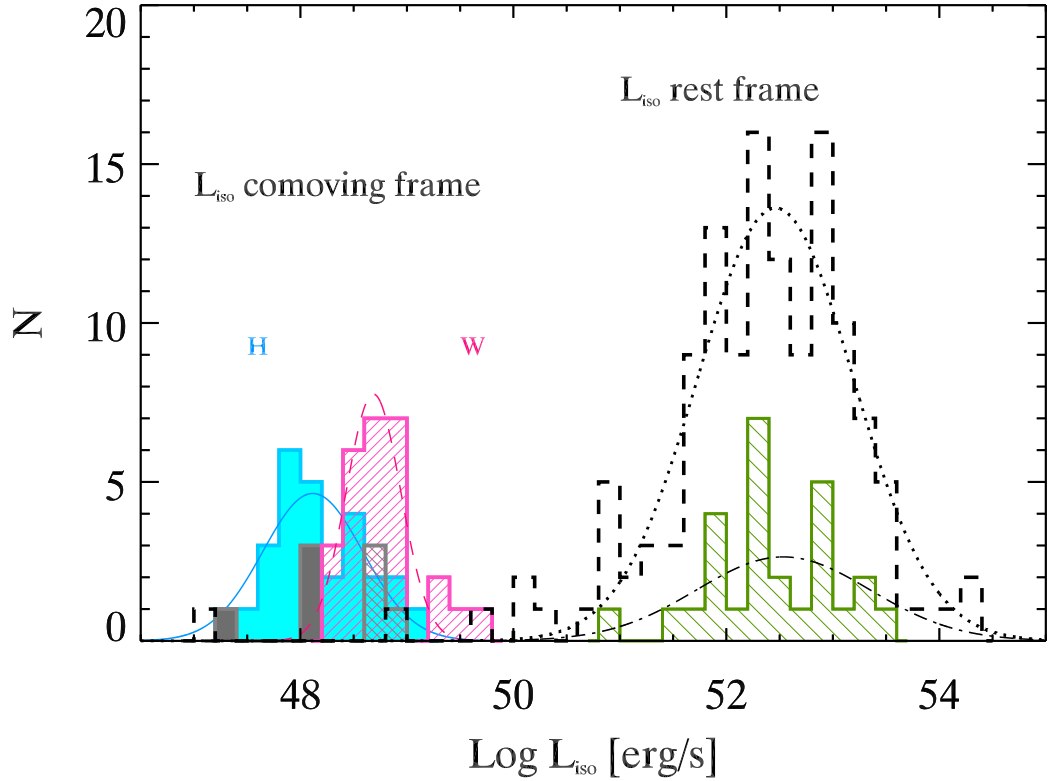,width=17cm}
\vskip -0.3 cm
\caption{
Isotropic luminosity distributions in the rest frame (dashed histogram) for the sample of 
131 GRBs with known redshift and constrained \epo. The hatched histogram shows the 30 GRBs of 
our sample for which we have an estimate of the peak of the afterglow. 
The beaming corrected distribution of \lisocom\ is shown by the solid filled histogram and 
hatched purple histogram for the H and W case 
for the 27 GRBs with a peak in the optical light curve. The four bursts with a peak in the GeV light curve are shown for 
comparison with the hatched and filled grey histograms.
}
\label{fg5}
\end{figure*}

In Fig. \ref{fg3}, \ref{fg4} and \ref{fg5} we show the distributions of the comoving 
frame peak energy, isotropic equivalent energy and luminosity. 
In Fig. \ref{fg3} we show the distributions of the peak energy: 
the sample of 132 GRBs with measured redshifts and known \ep\ is shown with 
the dashed line
and the subsample of 30 GRBs of this work for which we 
could estimate \G\  is shown with the red hatched histograms. 
These distributions represent \ep, i.e. the peak energy in the rest frame of the sources. 

The distributions of the comoving peak energy [derived as \epcom=\ep/(5\G/3)] are shown by the  
(cyan) filled and hatched (purple) histograms in Fig. \ref{fg3} for the H and 
W case, respectively,  
considering the 27 GRBs which show a peak in the optical light curve. 
Fig. \ref{fg3} shows also the fits with Gaussian functions: 
their parameters are reported in Tab. \ref{tab2}.

There is a reduction of the dispersion of the distribution of the peak energy from the 
rest frame to the comoving one. 
In the comoving frame \epcom\ clusters around $\sim$6 keV and $\sim$3 keV in the H and W case, 
respectively, with dispersions of nearly one decade, i.e. narrower than the dispersion of \ep.

Fig. \ref{fg4} shows the distribution of the isotropic energy \eiso\ for all the 132 GRBs with 
known $z$ and measured \ep\ (dashed line) and for the 30 GRBs with an estimate of 
\G\ (hatched red histogram).  
The \eisocom=\eiso/\G\ distributions are shown with the 
solid filled (cyan) histogram and the hatched (purple) histogram for the H and W case. 
These distributions are obtained with the 27 GRBs with a peak in the optical light curve. The three GRBs with 
a peak in the GeV light curve are only shown for comparison (hatched and filled gray histogram).
The distributions of \eisocom\ are wide. 
On average the comoving frame \eisocom$\sim$1--3$\times 10^{51}$ erg in both the H and W case, but 
there is a reduction of the dispersion of the distribution of \eiso\ from the rest ($\sigma_{\rm sc}=0.64$)
to the comoving frame ($\sigma_{\rm sc}=0.43$ and $\sigma_{\rm sc}=0.49$) for the W and the H case, respectively (see Tab. \ref{tab2}).

Finally, in Fig. \ref{fg5} we show the distribution of \liso\ for the 131 GRBs in the sample 
(dashed line), the distribution of \liso\ 
for the 30 GRBs with estimated \G\ (red hatched histogram) and the comoving frame 
\lisocom=\liso/$(4\Gamma^{2}_{0}/3)$ distribution (solid filled  cyan and hatched purple histograms 
for the H and W case, respectively,  obtained with the 27 GRBs with a peak in the optical light curve). 
Interestingly, we find a strong clustering of the comoving frame distribution of \lisocom. 
For the H case we find (see Tab. \ref{tab2} for the values of the Gaussian fits) an average 
\lisocom$\sim10^{48}$ erg s$^{-1}$ with a small dispersion (0.47 dex), while when using the \G\ 
computed in the wind density profile (W) case we find an almost universal value of 
\lisocom$\sim 5\times10^{48}$ erg s$^{-1}$ with a dispersion of less than one order of magnitude 
around this value (hatched purple histogram and dashed purple line in Fig. \ref{fg5}).  

\subsection{Comoving frame \amacom\ and \yonecom\ correlations}

\begin{figure*}
\hskip 0.9cm
\psfig{file=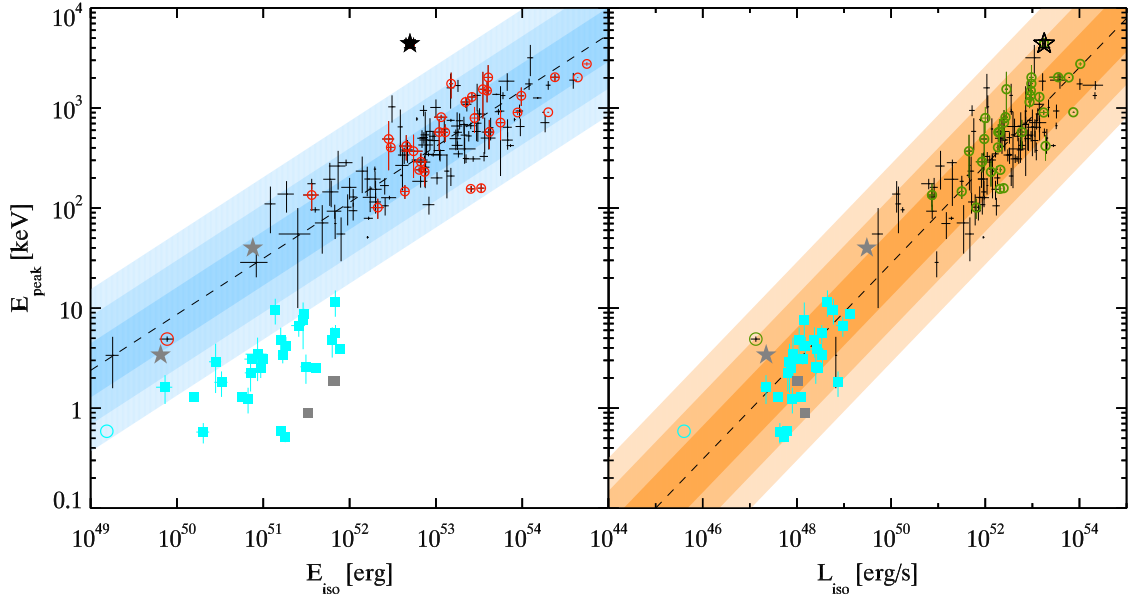,height=8cm}
\caption{
Homogeneous interstellar medium -- H.  
Left: \ama\ correlation in the rest frame (crosses and red circles) for 132 GRBs with $z$ 
and fitted \ep\ updated to May 2011. Right: \yone\ correlation with 131 GRBs. 
In both panels the best fit correlation is shown by the dashed line and its 1, 2, 3$\sigma$ scatter is 
shown by the shaded region. 
The comoving frame \epcom\ and \eisocom\ (left) and \epcom\ and 
\lisocom\ (right) of 30 GRBs (red open circles [left panel] and green open circles [right panel]) 
in our sample (Tab. \ref{tab1}) with an estimate of the \G\ factor  are shown with the filled cyan square symbols 
(27 events with $t_{\rm p,z}$ in the optical light curve) or grey filled square (the three long GRBs with a peak in the 
GeV light curve). 
The short GRB 090510 is also 
shown with a star symbol and the low luminosity GRB 060218 (with \G$\sim$5 [Ghisellini et al. 2006]) 
is shown with an open circle. 
}
\label{fg6}
\end{figure*}

Here we show the effect of correcting the spectral energy correlations 
\ama\ and \yone\ for  the bulk Lorentz factors \G. 
These correlations were originally found with a dozen of GRBs (Amati et al. 2002 
and Yonetoku et al. 2004 for the \ama\ and \yone\ correlations respectively) and since then 
updated with newly discovered GRBs with measured redshifts $z$ and well constrained 
spectral peak energies \ep. 
In this work we have updated the sample of GRBs with all these observables to May 2011. 
We have 132 GRBs with measured $z$ and known \ep\ and \eiso\ and 131 GRBs with 
measured $z$ and \ep\ and \liso. 
We show the corresponding \ama\ and \yone\ correlations in Fig. \ref{fg6} (left and right 
panel respectively). 
The best fit correlation parameters (obtained with the bisector method) are reported in 
Tab. \ref{tab4}. 
We find that $E_{\rm peak}\propto E_{\rm iso}^{0.56}$ (dashed line in Fig. \ref{fg6}) with a 
scatter $\sigma=0.24$ (computed perpendicular to the best fitting line and modeled with a 
Gaussian function). 
The other correlation is $E_{\rm peak}\propto L_{\rm iso}^{0.50}$ with a slightly larger 
scatter $\sigma=0.3$. 
The 1, 2 and 3$\sigma$ dispersion of the correlations are shown with the shaded stripes.

\begin{figure*}
\hskip 0.9cm
\psfig{file=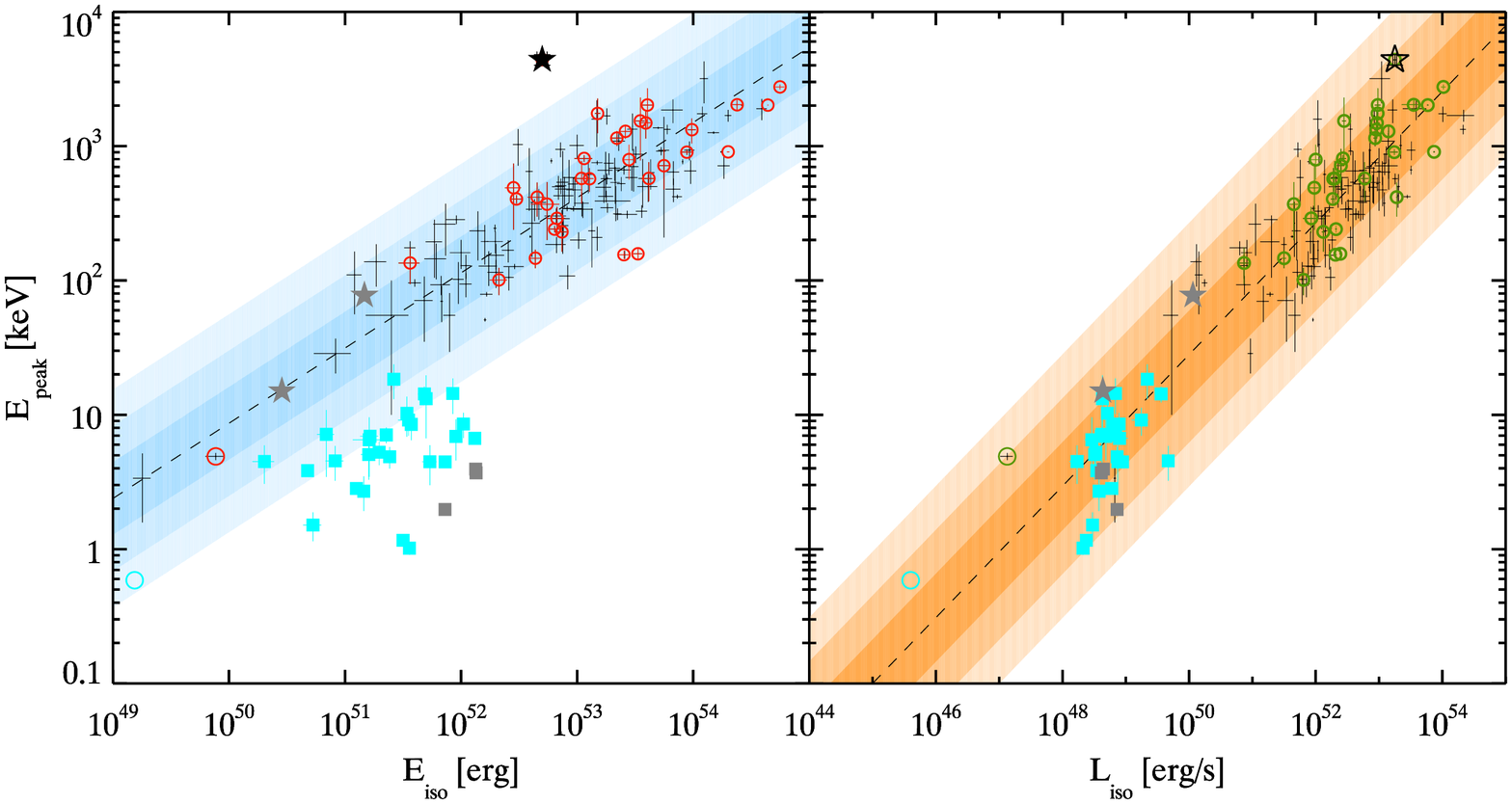,height=8cm}
\caption{
Wind interstellar medium -- W. Same as Fig. \ref{fg6}.}
\label{fg7}
\end{figure*}

Fig. \ref{fg6} also shows the comoving frame \epcom\ and \eisocom\ (left panel) 
and \epcom\ and \lisocom\ (right panel) for the 30 GRBs 
of our sample with an estimate of \G\ in the H case.  
The 27 GRBs with a peak in the optical are shown with the cyan filled squares in Fig.\ref{fg6} 
while the three long GRBs with a peak in the GeV light curve are shown with the filled gray squares.
Fig. \ref{fg7} show the same correlations (\ama\ and \yone\ in the left 
and right panels respectively) 
for the W case.
We note that in both the H and W cases there is a clustering of the points around typical 
values of \epcom, \eisocom\ and \lisocom. 
Tab. \ref{tab4} reports the correlation analysis among the comoving frame quantities.

\begin{table*}
\begin{center}
  \begin{tabular}{llllllll}
    \hline\hline
              &   Correlation      & \# GRBs     & $\rho$ & $P_{\rm chance}$ & $m$ & $q$    & $\sigma_{\rm sc}$ \\
    \hline
		&    \ama\             & 132           &   0.8     & $10^{-30}$   & 0.56$\pm$0.02   & -26.06$\pm$1.14 & 0.24 \\ 
		&    \ama\             & 27             &   0.71   & $3\times10^{-5}$   & 0.67$\pm$0.10   & -33.88$\pm$5.0 & 0.28  \\ 
		&    \ama\             & 30             &   0.76   & $10^{-6}$   			& 0.58$\pm$0.07   & -28.26$\pm$3.74 & 0.29  \\ 		
		&    \yone\             & 131          &   0.77     & $3\times10^{-26}$      & 0.49$\pm$0.04   & -23.03$\pm$1.84   & 0.30 \\ 
		&    \yone\             & 27            &   0.76   & $3\times10^{-6}$   & 0.65$\pm$0.08   & -31.53$\pm$4.36   & 0.25  \\ 
		&    \yone\             & 30            &   0.8   & $10^{-7}$                 & 0.57$\pm$0.06   & -27.14$\pm$3.37   & 0.27  \\ 

\hline
       Density   &   Correlation    & \# GRBs   & $\rho$ & $P_{\rm chance}$ &  &   &   \\
\hline
	H           &    \epcom--\eisocom & 27 & 0.62 & $6\times10^{-4}$ &   &   &  \\ 
		        &    \epcom--\eisocom & 30 & 0.43 & $2\times10^{-2}$ &   &   &  \\ 
		        &    \epcom--\lisocom  & 27 & 0.72 & $2\times10^{-5}$ &   &   &  \\ 
		        &    \epcom--\lisocom  & 30 & 0.68 & $3\times10^{-5}$ &   &   &  \\ 

		        &  &  &  &  &  &  &  \\ 		        
	W          &    \epcom--\eisocom & 27 & 0.41 & $4\times10^{-2}$ &   &   &  \\ 
		        &    \epcom--\eisocom & 30 & 0.28 & $0.3$                  &   &   &  \\ 
		        &    \epcom--\lisocom & 27 & 0.50 & $7\times10^{-3}$ &   &   &  \\ 
		        &    \epcom--\lisocom & 30 & 0.47 & $10^{-2}$ 		&   &   &  \\ 
\hline
     \hline
      \end{tabular}
\caption{
Results of the fit of the \ama\ and \yone\ correlations updated in this paper to May 2011. 
The Spearman correlation coefficient $\rho$ and the chance probability 
$P_{\rm chance}$ is given with the slope $m$ and normalization $q$ of the least square fits.} 
\label{tab4}
\end{center}
\end{table*}



\section{Discussion and Conclusions}

We have considered all bursts with measured $E_{\rm peak}$ and known redshift up to May 2011 (132 GRBs). 
Among these we have searched in the literature for any indication of the peak of the afterglow light curve $t_{\rm p,z}$ suitable 
to estimate the initial bulk Lorentz factor \G. 
Our sample of bursts is composed by  27 GRBs with a clear evidence of $t_{\rm p,z}$  in the optical light curve. 
We have derived the peak energy $E^\prime_{\rm peak}$, the isotropic energy 
$E^\prime_{\rm iso}$ and the isotropic peak luminosity $L^\prime_{\rm iso}$ in the comoving frame. 
To this aim we have derived the general formula for the computation of \G\ (\S.3) considering two possible scenarios:  
a uniform interstellar medium density profile ($n=$const, H) or a wind density profile ($n \propto r^{-2}$, W).
 
For the wind case the \G\--distribution (Fig. \ref{fg1} and Tab. \ref{tab2}) is shifted at somewhat smaller  
values ($\langle\Gamma_{0}\rangle\sim$ 66) 
than the same distribution for the homogeneous density case ($\langle\Gamma_{0}\rangle\sim$ 138). 
The distribution of $E^\prime_{\rm peak}$ is relatively narrow and centered around $\sim$6 keV 
 or $\sim 3$ keV for the W and H case (Fig. \ref{fg3} and Tab. \ref{tab2}). 
The distribution of $L^\prime_{\rm iso}$ (Fig. \ref{fg5}) clusters, especially for the wind case, 
in a very narrow range (much less than a decade), around $5\times10^{48}$ erg s$^{-1}$, 
while the distribution of $E^\prime_{\rm iso}$ (Fig. \ref{fg4}) 
is broader  and centered at $3\times10^{51}$ erg. 
$E_{\rm iso}$ and $L_{\rm iso}$ correlate with \G, ($\propto$\G$^{2.2}$
both for the wind and the homogeneous case) and the correlation is stronger 
(with a scatter $\sigma=0.07$) for the wind case.
Finally, the duration of the burst, as expected, does not correlate with \G.
 
The correlations that we have found are strong despite they are defined with a still small number of GRBs. We 
expect that with the increase of the number of GRBs with measured $t_{p,z}$ and well determined spectral properties (i.e. 
\ep, \eiso\ and \liso) the slope and normalization of these correlations might change. 
 
For comparison we also considered four GRBs with a peak in the GeV light curve. 
If the GeV emission is interpreted as afterglow (Barniol--Duran \& Kumar 2009; Ghirlanda et al. 2010; Ghisellini et al. 2010)
the measure of $t_{\rm p,z}$ at early times in the GeV range allows us to estimate their \G, that are consistent
with the correlations found using only the bursts with $t_{\rm p,z}$ observed in the optical.
Although not a proof, this is a hint in favour of the afterglow origin of the GeV emission.

These results are schematically summarized in the first column of Tab. \ref{tab5}.
The second column of the same table reports some immediate implications of these results.
Since $E^\prime_{\rm peak}\propto E_{\rm peak}$\G\ is contained in a narrow range,
all bursts emit their radiation at a characteristic frequency in their comoving frame,
irrespective of their bulk Lorentz factor.
Furthermore, we can assume that $E_{\rm peak}\propto$\G,
and this, together with the quadratic dependence on \G\ of
$E_{\rm iso}$ and $L_{\rm iso}$, yields the ``Amati" and
the ``Yonetoku" relations. {\it They are the result of a different \G--factors.} Indeed, at the 
extremes of the \ama\ and \yone\ correlations we find GRB 060218 which has the lowest \G$\sim 5$ (inferred from its
X--ray and optical properties -- Ghisellini, Ghirlanda \& Tavecchio  2007), while at the upper end (corresponding to 
the largest peak energies and isotropic energetics and luminosities) there is GRB 080916C which has the largest \G=880. 
The fact that the \ama\ and \yone\ correlations could be a sequence of \G\ factors has been also proposed by Dado, Dar \& De Rujula (2007) based on 
different assumptions. 

If all bursts had the same jet opening angle, then $L^\prime_\gamma =\theta_{\rm j}^2 L^\prime_{\rm iso}$,
and the (logarithmic) width of the $L^\prime_{\rm iso}$ distribution would be the same of the
(more fundamental) $L^\prime_\gamma$ distribution. On the other hand, we have some hints that very 
energetic and luminous GRBs tend to have narrower opening angles (e.g. Firmani et al. 2005).
It is this property that makes the collimation corrected $E_\gamma$ and $L_\gamma$ quantities
to correlate with $E_{\rm peak}$ in a different way (i.e. different slope) than in the Amati and 
Yonetoku relation (Ghirlanda et al. 2004; Nava et al. 2006).

We are then led to propose the following ansatz: the opening angle of the jet inversely correlates 
with the bulk Lorentz factor $\theta_{\rm j}\propto$ \G$^{-a}$.
There are too few GRBs in our sample with measured $\theta_{\rm j}$ to find
a reasonable value for the exponent $a$, but it is nevertheless
instructive to explore the case $a=1/2$, leading to $\theta^2_{\rm j}$\G$=$ constant.
If we assume this relation we find, for
the collimation corrected $E_{\gamma}$:
\begin{equation}
E_{\gamma} = \theta^2_{\rm j} E_{\rm iso} \propto \Gamma_{0} \propto E_{\rm peak}
\end{equation}
This is the ``Ghirlanda" relation in the wind case (Nava et al. 2006).
Similarly, for the collimation corrected luminosity (Ghirlanda, Ghisellini \& Firmani 2006):
\begin{equation}
L_{\gamma} = \theta^2_{\rm j} L_{\rm iso} \propto \Gamma_{0} \propto E_{\rm peak}
\end{equation}
Another important consequence of our ansatz is that, in the comoving frame,
the collimation corrected energetic $E^\prime_\gamma$ becomes constant:
\begin{equation}
E^\prime_{\gamma} = \theta^2_{\rm j} {E_{\rm iso}\over \Gamma_{0}}=\,\,{\rm constant} 
\end{equation}
This allows to ``re--intepret" the constancy of $L^\prime_{\rm iso}$ as a consequence
of the constant $E^\prime_\gamma$:
\begin{equation}
L^\prime_{\rm iso} \sim {E^\prime_\gamma \over T^\prime_{90} \theta^2_{\rm j}} = 
{E^\prime_\gamma \over T_{90}  \theta^2_{\rm j}\Gamma_0} =
\,\,{\rm constant} 
\end{equation}
In other words, in the comoving frame, the burst emits {\it the same amount of energy at the
same peak frequency}, irrespective of the bulk Lorentz factor.
For larger \G\ the emitting time in the comoving frame is longer
(by a factor \G\ if the observed $T_{90}$ is the same),
so the comoving luminosity is smaller. But since the jet opening
angle is also smaller (for larger \G), the isotropic equivalent
luminosity turns out to be the same. 
These consequences are listed in the third column of Tab. \ref{tab5}.

Interestingly, we note that the general formula for the estimate of the jet opening angle
\begin{equation}
\theta_{\rm j}\propto \left(\frac{t_{\rm j,obs}}{1+z}\right)^{{3-s}\over{8-2s}} 
\left(\frac{n_{0}\eta}{E_{\rm iso}}\right)^{{1}\over{8-2s}}
\end{equation}
with $s=0$ for the homogeneous case and $s=2$ for the wind case, can be combined 
with Eq. \ref{eqgamma} to give:
\begin{equation}
\theta_{\rm j}\Gamma_{0}\propto 
\left(\frac{t_{\rm j,obs}}{t_{\rm p,obs}}\right)^{{3-s}\over{8-2s}} 
\end{equation}
The product $\theta{\rm j}\Gamma_{0}$ then depends only on two observables, 
i.e. the time of the peak of the afterglow $t_{\rm p,obs}$ and 
the time of the jet break $t_{\rm j,obs}$, and it is independent from the 
redshift $z$ and the energetic \eiso\ as well as from the density profile normalization
$n_{0}$ and radiative efficiency $\eta$. If also the product 
$\theta_{\rm j}^2\Gamma_{0}=$const, then we can derive both 
$\theta_{\rm j}\propto (t_{\rm p,obs}/t_{\rm j,obs})^{{3-s}\over{8-2s}}$ 
and $\Gamma_{0}\propto (t_{\rm j,obs}/t_{\rm p,obs})^{{3-s}\over{4-s}}$. 
If the ansatz $\theta_{\rm j}^2\Gamma_0=$ const will prove to be true,
then by simply measuring the peak time and the jet break time of the afterglow light curve
we could estimate both $\theta_{\rm j}$ and $\Gamma_{0}$ for any GRB.

\begin{table*}
\begin{center}
\begin{tabular}{llll}
\hline\hline
Our results                        &Implications                                                        &If $\theta_{\rm j}^2\Gamma\sim$const        \\
\hline
$E^\prime_{\rm peak}\sim$ const &$E_{\rm peak} \propto\Gamma$                                           &    \\ 
$E_{\rm iso} \propto \Gamma^2$  &$E_{\rm iso}\propto E^2_{\rm peak}$                                    &$E_\gamma = \theta_{\rm j}^2 E_{\rm iso}\propto \Gamma\propto E_{\rm peak}$ \\ 
$L_{\rm iso} \propto \Gamma^2$  &$L_{\rm iso}\propto E^2_{\rm peak}$                                    &$L_\gamma= \theta_{\rm j}^2 L_{\rm iso}\propto \Gamma\propto E_{\rm peak}$ \\
$T_{90}$ not $f(\Gamma)$                  &$T^\prime_{90} \propto \Gamma$                                             &$E^\prime_\gamma \sim $ const     \\
$L^\prime_{\rm iso}\sim$ const            &$E^\prime_{\rm iso}/L^\prime_{\rm iso}\propto T^\prime_{90}\propto \Gamma$ &$L^\prime_\gamma \sim E^\prime_\gamma/T^\prime_{90}\sim 1/\Gamma$     \\
\hline
\hline
\end{tabular}
\caption{Schematic summary of our results and their implications for the case of a wind density profile.
We have assumed that both $E_{\rm iso}$ and $L_{\rm iso}$ scale as $\Gamma^2$, instead of $\Gamma^{2.2}$.
} 
\label{tab5}
\end{center}
\end{table*}

\begin{figure}
\hskip -0.4true cm
\psfig{file=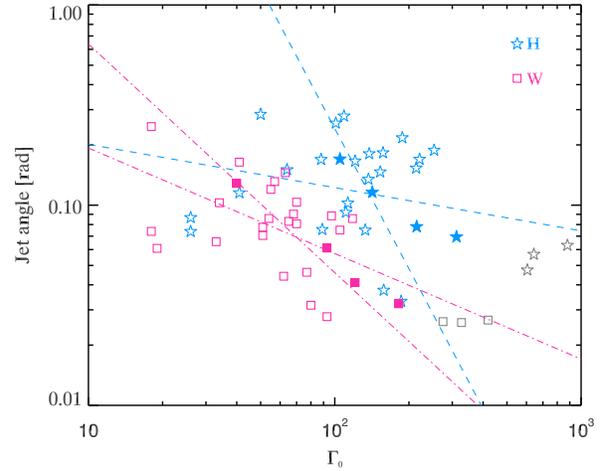,height=7cm}
\caption{
Jet opening angle as a function of $\Gamma_0$ for a H (stars) and for a W (squares). 
Empty symbols show the jet angles estimated by assuming the consistency of our sample with the 
$E_{\rm peak}$--$E_\gamma$ relation. Filled symbols refer to the bursts of our sample for which 
the jet opening angle has been calculated from the measured jet break time of the optical light curves. 
The two lines (dashed for the H case and dot--dashed for the W case) show the powerlaw fit of the data points 
considering $\theta_{\rm jet}$ vs $\Gamma_{0}$ and $\Gamma_{0}$ vs $\theta_{\rm jet}$ . 
The gray symbols show the three long bursts with a peak in the GeV light curve that, if interpreted as afterglow emission, 
allows us to estimate \G.
}
\label{fg8}
\end{figure}
\begin{figure}
\vskip -0.5 cm
\hskip -0.8true cm
\psfig{file=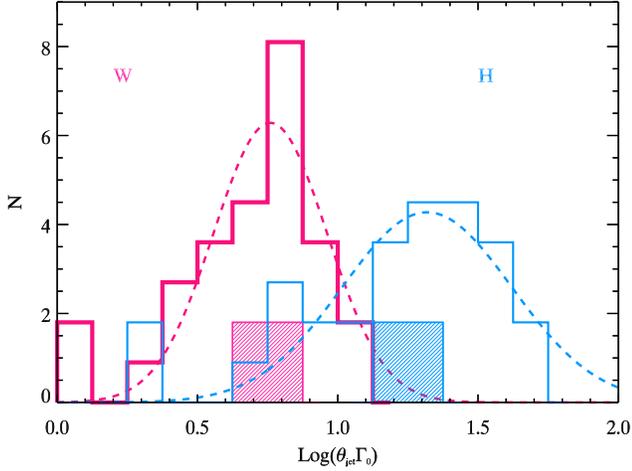,height=7cm}
\caption{
Distribution of $\theta_{\rm j}\Gamma_{0}$ in the H and W case (blue and purple histograms) 
estimated by assuming the $E_{\rm peak}$--$E_\gamma$ relation in the H (Ghirlanda et al. 2004) 
or W (Nava et al. 2006) case. The hatched histograms show the few GRBs in our samples for which 
$\theta_{\rm j}$ has been calculated from the measured jet break time in the optical light curve. }
\label{fg9}
\end{figure}

In our sample, only for 4 bursts we can estimate the jet opening angle from the measure of the jet 
break time of the optical light curve. 
Their small number does not make possible to directly test the 
existence of a relation between $\Gamma_0$ and $\theta_{\rm j}$. 
However, an estimate of the jet 
opening angle can be possible by assuming that all bursts in our sample are consistent with the 
``Ghirlanda" relation. 
Fig. \ref{fg8} shows the estimated $\theta_{\rm j}$ as a function of $\Gamma_0$. 
Stars (squares) refers to angles derived under the assumption of a H (W). To estimate the jet 
opening angles we considered the most updated ``Ghirlanda" correlation, which comprises 29 GRBs with 
measured jet break time (Ghirlanda et al. 2006). For the homogeneous density profile the relation has 
the form $\log E_{\rm peak}=-32.81+0.70 \log E_\gamma$, while in the case of a W the relation 
becomes $\log E_{\rm peak} =-50.08+1.04 \log E_\gamma$. 
Given the large scatter of the data points in Fig.~\ref{fg8}, we fitted both 
$\theta_{\rm j}$ versus $\Gamma_{0}$ and $\Gamma_{0}$ versus $\theta_{\rm j}$: we obtain 
$\theta_{\rm j}\propto \Gamma_0^{-0.22}$ and $\Gamma_0 \propto \theta_{\rm j}^{-2.32}$
for the H case (dashed lines in Fig. \ref{fg8}) and 
$\theta_{\rm j}\propto \Gamma_0^{-0.52}$ and $\Gamma_0 \propto \theta_{\rm j}^{-1.14}$
for the W case (dot--dashed line in Fig. \ref{fg8}). 
We conclude that our ansatz $\theta_{\rm j} \propto \Gamma_0^{-1/2}$ is consistent with, but not
proven by, this analysis.

An interesting exercise is to estimate the product $\theta_{\rm j}$\G. 
From the observational point of view $\theta_{\rm j}$\GA$\gg$1 at the end of the prompt phase, so 
that the decrease of \GA\ in the afterglow phase, due to the interaction of the GRB fireball with the 
interstellar medium, gives rise to a jet break when $\theta_{\rm j}$\GA$\sim$1. 

Some numerical simulations (Komissarov et al., 2009) of jet acceleration have shown that a magnetic 
dominated jet confined by an external medium should have $\theta_{\rm j}$\G$\le 1$. 
This value is inconsistent with typical values of $\theta_{\rm j}$ and \G: in the case of 
an homogeneous wind density profile the typical $\theta_{\rm j}\sim 0.1$ radiants (Ghirlanda et al. 2007) 
while in the case of a wind density profile $\theta_{\rm j}\sim 0.07$ radiants. 
Combining these values with the average values of \G\ estimated in this paper 
(Tab. \ref{tab1}) we find $\theta_{\rm j}$\G$\sim 14$ (5) for the H (W) case. 

These are approximate values: the sample of GRBs with measured $\theta_{\rm j}$ 
(Ghirlanda et al. 2007) contains only 4 bursts of the 
sample of  events of the present paper with estimated \G. 
However, though somehow speculative, 
we can derive $\theta_{\rm j}$ for the 32 GRBs of our sample assuming the \ghi\ correlation in the 
H case (Ghirlanda et al. 2004) or in the W (Nava et al. 2006). 
In Fig. \ref{fg9} we show the distributions of the product $\theta_{\rm j}$\G\ in the 
H case (blue histogram) and in the W case (purple histogram). 
We note that both are centered around typical values of 20 and 6 (for the H and W case, respectively). 
These values are in good agreement with the results of recent simulations of 
(i) a magnetized jet confined by the stellar material that freely expands when it 
breaks out the star (Komissarov, Vlahakis \& Koenigl  2010) or 
(ii) a magnetized unconfined split--monopole jet (Tchekhovskoy, McKinney \& Narayan 2009; 
Tchekhovskoy, Narayan \& McKinney 2010). 
A possible test of these two scenarios could be short GRBs where the absence of the progenitor 
star would prefer model (ii) for the jet acceleration. 
In our sample only the short/hard GRB 090510 is present. 
No jet break was observed for this event and in general we do not yet know if 
short GRBs follow the same \ghi\ correlation of long ones.

\section*{Acknowledgments}
We acknowledge ASI (I/088/06/0) and a 2010 PRIN--INAF grant for
financial support.  
We acknowledge the referee for comments and suggestions that improved this work.


\begin{thebibliography}{}
\bibitem{} Abdo, A. A., Ackermann, M., Ajello, M., et al., 2009, Nature, 462, 331
\bibitem{} Abdo A. A,  Ackermann M., Ajello M., et al., 2009a, ApJ, 706, L138
\bibitem{} Ackermann M., Asano K., Atwood W.B. et al.,  2010, ApJ, 716, 1178
\bibitem{} Amati L., Frontera F., Tavani M. et al., 2002, A\&A, 390, 81
\bibitem{} Amati L., Frontera F. \& Guidorzi C., 2009, A\&A, 508, 173
\bibitem{} Band D.L. \& Preece R., 2005, ApJ, 627, 319
\bibitem{} Beloborodov A.M., 2002, ApJ, 565, 808
\bibitem{} Bianco C. L. \& Ruffini R., 2005, ApJ, 633, L13
\bibitem{} Blandford R.D. \& McKee C.F., 1976, Phys. of Fluids, 19, 1130 
\bibitem{} Bosnjak Z., Celotti A., Longo F. et al., 2008, 384, 599
\bibitem{} Butler N.R., Kocevski D., Bloom J.S. \& Curtis J.L., 2007, ApJ, 671, 656
\bibitem{} Butler N.R., Kocevski D. \& Bloom J.S.,  2009, ApJ, 694, 76
\bibitem{} Chevalier R., Li W., 1999, ApJ, 520, L29
\bibitem{} Costa E., Frontera F., Heise J. et al., 1997, Nature, 387, 783
\bibitem{}  Dado S., Dar A. \& De Rujula A., 2007, ApJ, 663, 400
\bibitem{} De Pasquale, M., Schady, P., Kuin, N. P. M., et al., 2010, ApJ, 709, L146
\bibitem{} Firmani, C., Ghisellini, G., Ghirlanda, G., et al., 2005, MNRAS, 360, L1
\bibitem{} Frail D., Kulkarni S. R.; Nicastro L., et al., 1997, Nature, 389, 261
\bibitem{} Ghirlanda G., Ghisellini G., Lazzati D., 2004, ApJ, 616, 331
\bibitem{} Ghirlanda G., Ghisellini G., Firmani C., Celotti A. \& Bosnjak Z., 2005, MNRAS, 360, 45
\bibitem{} Ghirlanda G., Ghisellini G., Firmani C., 2006, NJPh, 8 123
\bibitem{} Ghirlanda G., Nava L., Ghisellini G., 2010, A\&A, 511, 43
\bibitem{} Ghirlanda G., Ghisellini G., Nava L., 2010a, A\&A, 510, L7
\bibitem{} Ghirlanda G., Ghisellini G., Nava L., Burlon, D., 2011, MNRAS, 401, L47
\bibitem{} Ghisellini G., Ghirlanda G., Tavecchio F., 2007, MNRAS, 382, L77
\bibitem{} Ghirlanda G., Nava L., Ghisellini G., Firmani C. \& Cabrera J.I., 2008, MNRAS, 387, 319
\bibitem{} Ghisellini G., Nardini M., Ghirlanda G., Celotti A., 2009, MNRAS, 393, 253
\bibitem{} Ghisellini G., Ghirlanda G., Nava L., Celotti A., 2010, MNRAS, 403, 926
\bibitem{} Gruber, D., Kruehler, T., Foley, S., et al., 2011, A\&A, 528, 15
\bibitem{} Hasco't R., Vennin V., Daigne F., Mochkovitch R., 2011, arXiv1101.3889
\bibitem{} Isobe T., Feigelson E. D., Akritas M. G., Babu G. J., 1990, ApJ, 364, 104
\bibitem{} Komissarov S. S., Vlahakis N., Koenigl A., Barkov, Maxim V., 2009, MNRAS, 394, 1182
\bibitem{} Komissarov S. S., Vlahakis N., Koenigl A., 2010, MNRAS, 407, 17
\bibitem{} Krimm H.A., Yamaoka K., Sugita S. et al., 2009, ApJ, 704, 1405
\bibitem{} Liang E.-W.; Yi S.-X.; Zhang, J., 2010, ApJ, 725, 2209
\bibitem{} Lithwick Y.; Sari R., 2001, ApJ, 555, 540
\bibitem{} Lv J.,  Zou Y.-C.,  Lei W.-H., 2011, arXiv:1109.3757
\bibitem{} Molinari E.,  Vergani, S. D., Malesani, D., et al., 2007, A\&A, 469, L13
\bibitem{} Nakar, E. \& Piran, T., 2005, MNRAS, 360, L73
\bibitem{} Nava L., Ghisellini G., Ghirlanda G., et al., 2006, 450, 471
\bibitem{} Nava L., Ghirlanda G., Ghisellini G. \& Firmani C., 2008, MNRAS, 391, 639
\bibitem{} Panaitescu A., Kumar P., 2000, ApJ, 543, 66
\bibitem{} Piran, T. \& Nakar, E., 2010, ApJ, 718, L63
\bibitem{} Sari R., 1997, ApJ, 489, L37
\bibitem{} Sari R., Piran T., 1999, ApJ, 520, L17
\bibitem{} Shahmoradi, A. \& Nemiroff, R. J., 2011, MNRAS, 411, 1843
\bibitem{} Tchekhovskoy A., McKinney J. C., Narayan R., 2009, ApJ, 699, 1789
\bibitem{} Tchekhovskoy A., Narayan R., McKinney J. C., NewA, 2010, 15, 749
\bibitem{} Yonetoku, D., Murakami, T., Nakamura, T. et al. 2004, ApJ, 609, 935
\bibitem{} Wijers R. A. M. J. \& Galama T. J., 1999, ApJ, 523, 177
\bibitem{} Zhao X.-H., Li Z., Bai J.-M., 2011, ApJ, 726, 89
\bibitem{} Zou Y.-C., Piran T., et al., 2010, MNRAS, 402, 1854
\bibitem{} Zou, Y.-C., Fan, Y.-Z., Piran T., 2011, ApJ, 726, L2	
\end{thebibliography}
\end{document}